\documentclass[useAMS]{mn2e} 
\usepackage{times}
\usepackage{psfig} \usepackage{epsfig} 
\usepackage{amssymb}
\usepackage{amsbsy}
\def\etal{{\it et al.\ }} 
\def\km{{\rm\thinspace km}} \def\kmps{${\rm\thinspace
km}{\rm\thinspace s}^{-1}$}   \def\omegam{{\Omega_{\rm m}}}
\def\beqn{\vspace{2mm} \begin{eqnarray}} \def\eeqn{\vspace{2mm}
\end{eqnarray}} \def\beq{\vspace{2mm} \begin{equation}}
\def\eeq{\vspace{2mm} \end{equation}}   \def\beqn{\vspace{2mm}
\begin{eqnarray}} \def\eeqn{\vspace{2mm} \end{eqnarray}}
\def\km{{\rm\thinspace km}} \def\s{{\rm\thinspace s}}
\def\kmps{\hbox{$\km\s^{-1}\,$}} 

\def\gsim{\lower.73ex\hbox{$\sim$}\llap{\raise.4ex\hbox{$>$}}$\,$}
\def\lsim{\lower.73ex\hbox{$\sim$}\llap{\raise.4ex\hbox{$<$}}}
\def\ss{{\scriptscriptstyle S}} \def\rr{{\scriptscriptstyle R}}
\def\np{{n^\prime}} \def\npp{{n^{\prime\prime}}}

\begin{document}

\title{Reconstructed Density and Velocity Fields from the 2MASS Redshift
Survey}

\author[Erdo\u{g}du \etal] {
\parbox[t]{\textwidth}{
Pirin Erdo\u{g}du$^{1,2,3}$, 
Ofer Lahav$^{3,4}$,
John P. Huchra$^{5\star}$,
Matthew Colless$^{6}$, 
Roc M. Cutri$^{7}$, 
Emilio Falco$^{5 \star}$
Teddy George$^{6}$,
Thomas Jarrett$^{8}$, 
D. Heath Jones$^{7}$, 
Lucas M. Macri$^{9}$, 
Jeff Mader$^{10}$,
Nathalie Martimbeau$^{5}$, 
Michael A. Pahre$^{5}$, 
Quentin A. Parker$^{7,11}$,
Ana\"is Rassat$^{4}$, 
Will Saunders$^{7}$ }
\vspace*{6pt} \\ 
$^{1}$  Department of Physics, Middle East Technical University, 06531, Ankara, Turkey \\ 
$^{2}$  School of Physics \& Astronomy, University of Nottingham, University Park, Nottingham, NG7 2RD, UK\\
$^{3}$  Institute of Astronomy, Madingley Road, Cambridge, CB3 0HA, UK\\
$^{4}$  Department of Physics and Astronomy, University College London, Gower Street, London WC1E 6BT, UK\\
$^{5}$  Harvard-Smithsonian Centre for Astrophysics, 60 Garden Street, MS-20, Cambridge, MA 02138, USA\\ 
$^{6}$  Canada-France-Hawaii Telescope Corporation, 65-1238 Mamalahoa Hwy Kamuela, Hawaii 96743, USA\\
$^{7}$  Anglo-Australian Observatory, PO Box 296, Epping,NSW 2052, Australia\\ 
$^{8}$  Infrared Processing and Analysis Center, California Institute of Technology, Pasadena, CA 91125, USA\\ 
$^{9}$  National Optical Astronomy Observatory, 950 North Chrry Avenue, Tucson, AZ 85726, USA\\ 
$^{10}$ W.M. Keck Observatory, Kamuela, HI 96743, USA\\ 
$^{11}$ Department of Physics, Macquarie University, Sydney, NWS 2109, Australia\\
$^\star$ Guest Observer, Cerro Tololo Interamerican Observatory, operated by AURA for the National Science Foundation \\
}
\maketitle

\begin{abstract}
We present the reconstructed real-space density and the predicted velocity fields from the Two Mass
Redshift Survey (2MRS). The 2MRS is 
the densest all-sky redshift survey to date and 
includes about 23,200 galaxies with extinction corrected 
magnitudes brighter than $K_{\rm s}=11.25$.
Our method is based on the
expansion of these fields in Fourier-Bessel functions. 
Within this framework, the linear redshift distortions only 
affect the density field in the radial direction and can easily be deconvolved
using a distortion matrix. Moreover, in this coordinate system, 
the velocity field is related to the density field by a simple 
linear transformation. 
The shot noise errors in the reconstructions 
are suppressed by means of a Wiener filter which yields 
a minimum variance estimate of the density and velocity fields.
Using the reconstructed real-space density fields, we identify all major
superclusters and voids. 
At 50 $h^{-1}$ Mpc, 
our reconstructed velocity field indicates a back-side infall to the Great
Attractor region of $v_{\rm infall}=(491\pm200) (\beta/0.5) \kmps$ in the Local Group
frame and $v_{\rm infall}=(64\pm205) (\beta/0.5) \kmps$ in the cosmic microwave
background (CMB) frame and $\beta$ 
is the redshift distortion parameter. 
The direction of the reconstructed dipole agrees well
with the dipole derived by Erdo\u{g}du \etal (2006). The misalignment between
the reconstructed 2MRS and the CMB dipoles drops to 13$^\circ$ at around 5000 $\kmps$ but then increases at larger distances.
\end{abstract}

\section{Introduction}\label{sec:2mass:introduction}

A cosmographical description of the galaxy distribution is crucial to our understanding of the mechanisms 
of structure formation that generate the complex pattern of sheets and 
filaments comprising the `cosmic web'.
Today, there are many more redshifts available for galaxies than velocity measurements. This discrepancy has inspired a great deal of work on methods for the reconstruction of the mass and velocity fields 
from redshifts alone.
These methods use a variety of functional representations (e.g. Cartesian, 
Fourier, spherical harmonics or wavelets)
and smoothing techniques (e.g. a Gaussian or a top-hat sphere). However, several complications 
arise when one tries to reconstruct peculiar velocity and real-space mass density fields 
directly from redshift surveys. First, one must relate the light distribution to the mass field. On large scales, this relationship is generally assumed to be linear with a proportionality constant (but see e.g. Wild \etal 2005; Conway \etal 2005; Marinoni \etal 2005 who argue that biasing is non-linear and redshift dependent even on large scales). 
Another common assumption is that the galaxy distribution samples the underlying smooth
mass field and the
finite sampling of the smooth field introduces 
Poisson `shot noise' errors\footnote{This is only an approximation. A more
  realistic model for galaxy clustering is the halo model (e.g. Cooray \&
  Sheth 2002) where the linear bias parameter depends on the mass of the dark matter haloes where the galaxies reside. For this model, the mean number of galaxy pairs in a given halo is usually lower than the Poisson expectation.}.  
Yet another important physical problem associated with the 
recovery of the real-space mass density field from the galaxy density field is the correction of the distortions in the redshift space clustering pattern. 
Furthermore, in redshift surveys the actual number of galaxies in a
given volume is larger than the number observed, particularly in
f\mbox{}l\mbox{}ux limited samples where at large distances only the very
luminous galaxies can be seen. 

In this paper, we recover the
real-space density and predicted velocity fields of the 2MRS (Huchra \etal 2005).  The reconstruction procedure used is based on linear theory and 
closely follows Fisher \etal (1995, hereafter FLHLZ).  The density field in
redshift-space is decomposed into spherical harmonics and Bessel
functions ({\it Fourier-Bessel functions}). 
Then the real-space density and velocity fields are reconstructed from those 
in the redshift-space using a Wiener filter. 

The idea of cosmography using spherical harmonics goes back to Peebles
(1973) but the method only gained popularity with the advent of
all-sky surveys. Since then, it has been applied
to the Infrared Astronomical  Satellite ($IRAS$) 
selected galaxy surveys (e.g. to the $IRAS$ 1.2 Jy 
Survey by Fabbri \& Natale 1990; Scharf
\etal 1992; Scharf \& Lahav 1993; Fisher, Scharf \& Lahav 1994; Lahav
\etal 1994; Nusser \& Davis 1994; Heavens \& Taylor 1995 and to $IRAS$ 
Point Source Catalogue Redshift Survey  (PSCz) by Tadros \etal 1999, Hamilton, Tegmark \& Padmanabhan 2000; 
Taylor \etal 2001) and the 
peculiar velocity catalogues (e.g. to elliptical galaxy sample by 
Reg\"os \& Szalay 1989; to Mark III galaxy sample by Davis, Nusser \& Willick 1996 and 
Hoffman \etal 2001). The spherical harmonics technique has 
proven to be very convenient for addressing many of the problems
inherent to the analyses of redshift surveys. 
The spherical harmonics are one of 
the most convenient ways of smoothing the noisy data. 
A careful choice of
spherical coefficients greatly reduces the statistical
uncertainties and errors introduced by the non-linear effects.
Furthermore, the treatment of linear redshift distortions are
particularly straightforward in Fourier-Bessel space, especially for 
surveys which have almost all-sky coverage and dense sampling such as 2MRS. 
     
The Wiener-filtered fields are optimal reconstructions in the
sense that the variance between the derived reconstruction and the
underlying true field is minimised. As opposed to {\it ad hoc} 
smoothing schemes,
the smoothing due to the Wiener filter is determined by the data. In the limit 
of high
signal-to-noise, the Wiener filter modifies the observed data only
weakly, whereas it strongly suppresses the contribution of the
data contaminated by shot noise.
The three-dimensional Wiener reconstruction of cosmological
fields in Fourier-Bessel space was first developed by FLHLZ and
applied to the $IRAS$ 1.2 Jy survey. Webster \etal (1997, hereafter
WLF) expanded on the initial results of FLHLZ. 
Later, Schmoldt \etal (1999) applied the same method to the 
$IRAS$ PSCz survey to recover density and velocity fields. 

This paper is structured as follows: in the next section, we describe the
Two Micron Redshift Survey.
In Section~\ref{sec:2mass:sph}, we give a
brief summary of the decomposition of the density field in
redshift-space into a set of orthogonal spherical harmonics and Bessel
functions. In Section~\ref{sec:2mass:zdistort}, we discuss how these
redshift-space harmonics are related to those in real-space. In
Section~\ref{sec:2mass:wf}, we give expressions 
for the Wiener filter in Fourier-Bessel space. 
The implementation of the method to 2MRS is
given in Section~\ref{sec:2mass:app}, along with the maps of the
density and velocity fields.  In the same section, we also show
reconstruction of the Local Group (LG) acceleration 
determined directly from
harmonic coefficients. The last section contains
a discussion and summary of the work. Throughout this paper, we
assume that the galaxy distribution is related to the underlying smooth
mass density field by the linear bias parameter, $b$, so that
$\delta_g ({\bf x}) = b \delta_m ({\bf x})$. With this notation, the constant of proportionality 
between velocity and the density field is $\beta$, where $\beta=\omegam^{0.6}/b$. 
It is worth emphasising that one of the ultimate goals of this work is to test linear biasing 
by comparing the reconstructed velocity fields with the observed peculiar velocity field.  
The distances are given in units of velocity and the Hubble constant is 
$H_0$ = 100 $h$ ${\rm km} {\rm s}^{-1} {\rm Mpc}^{-1}$.

\section{The Two Micron All-Sky Redshift Survey}\label{sec:dip:data}

\begin{figure*}
\psfig{figure=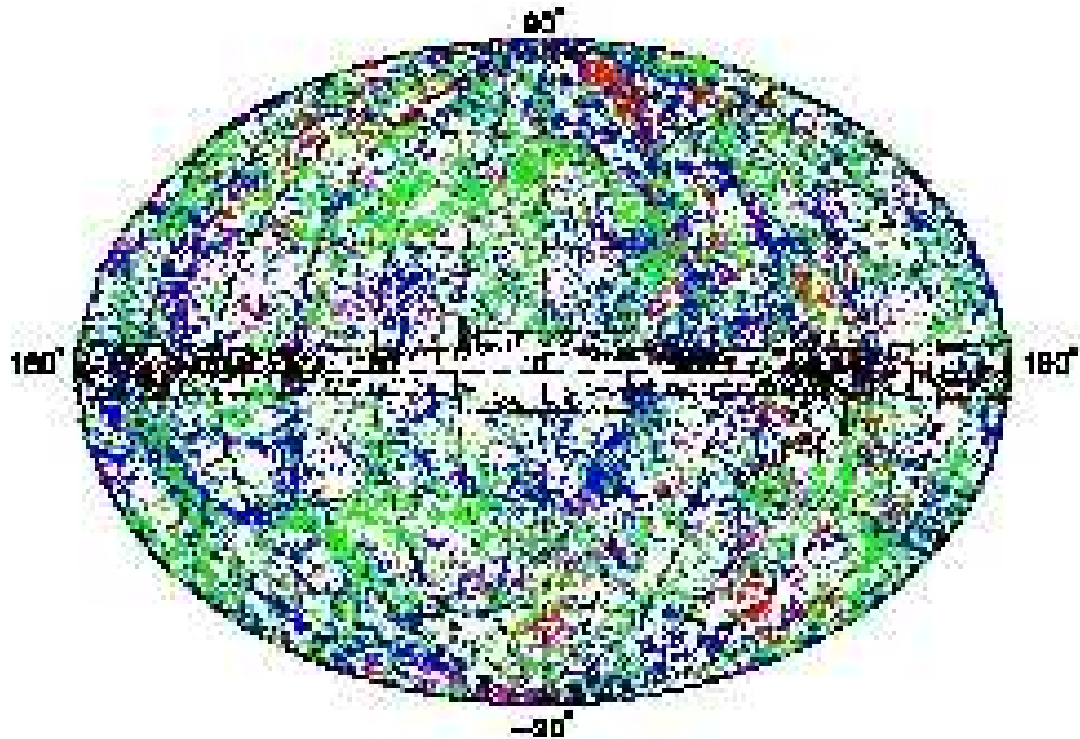,angle=0,width=\textwidth,clip=}
\caption[All Objects in the 2MASS Redshift Catalogue in Galactic
Aitoff Projection] {All Objects (24 788 galaxies) in the 2MASS Redshift Catalogue in
Galactic Aitoff Projection. The plot contains Galaxies with ${\rm z}\leq0.01$ 
plotted in red, $0.01<{\rm z}\le0.025$ are plotted in blue,
$0.025<{\rm z}<0.05$ are plotted in green and ${\rm z}\geq0.05$ are
plotted in magenta. Galaxies without measured redshifts are plotted in
black. The zone of avoidance is outlined by dashed lines.}
\label{fig:2MRS}
\end{figure*}

The Two Micron All-Sky Redshift Survey (2MRS) is the densest all-sky
redshift survey to date. The
galaxies in the northern celestial hemisphere are 
being observed by the Fred Lawrence Whipple Observatory (FLWO) 1.5-m telescope, 
the Arecibo 305-m telescope and the Green Bank 100-m Telescope.
In the southern hemisphere, most 
galaxies at high galactic latitude (about 6000 galaxies)
were observed as a part of the 6-degree field galaxy survey  
\footnote{6dFGS has a much fainter  
magnitude limit  ($K_{\rm s}=12.75$) than 2MRS and contains around 150,000
galaxy redshifts. The 6dFGS documentation and the database 
can be found on the WWW at 
http://www.aao.gov.au/local/www/6df/}
(6dFGS, Jones \etal 2004; Jones \etal 2005)
and the low galactic latitude galaxies are being observed at 
CTIO (by L. Macri and J.P. Huchra). 
The first phase of 2MRS is now
complete. In this phase we obtained redshifts for approximately
23,150 2MASS galaxies from a total sample of 24,773 galaxies with
extinction corrected magnitudes (Schlegel, Finkbeiner \& Davis 1998)
brighter than $K_{\rm s}=11.25$. This magnitude limit corresponds to a
median redshift of $z\approx0.02$ ($\approx 6000 \kmps$).  Almost all 
of the 1600 galaxies that remain without redshifts are at
very low galactic latitudes ($|b|\lesssim5^\circ$) 
or obscured/confused by the dust and the
high stellar density towards the Galactic
Centre. Figure~\ref{fig:2MRS} shows all the objects in 2MRS in
Galactic Aitoff Projection. Galaxies with ${\rm z}\leq0.01$ are
plotted in red, $0.01<{\rm z}\le0.025$ are plotted in blue,
$0.025<{\rm z}<0.05$ are plotted in green and ${\rm z}\geq0.05$ are
plotted in magenta. Galaxies without measured redshifts are plotted in
black. 2MRS can be compared with the deeper 2MASS galaxy catalogue
(K$<$14th mag) shown in Jarrett (2004, Figure 1).
\subsection{Survey Completeness}\label{sec:2masscomp}
2MASS\footnote{2MASS
database and the full documentation are available on the WWW at
http//www.ipac.caltech.edu/2mass.} 
has good photometric uniformity and an unprecedented
integral sky coverage. The photometric uniformity is better than
$4\%$ over the sky including the celestial poles (e.g. Jarrett \etal 2000$a$,
2003) and at magnitudes brighter than $K_{\rm s}=12$, 2MASS is essentially complete
 down to very low galactic latitudes (to $b^2\circ$, e.g. Huchra \etal 2005). 
The uniform completeness of the galaxy sample is slightly limited by the
presence of foreground stars, for a typical high latitude sky less
than $2\%$ of the area is masked. These missing regions are
accounted for using a coverage map, defined as the fraction of the
area of an 8\arcmin$\times$8\arcmin pixel that is not obscured by
stars brighter than 10th mag.  Galaxies are then weighted by the
inverse of the completeness, although the analysis is almost unaffected
by this process as the completeness ratio is very close to one for
most parts of the sky.

The stellar contamination of the catalogue is low and is reduced
further by manually inspecting the objects below a heliocentric redshift of
$cz=200$\kmps.  
The foreground stellar confusion is highest at low
Galactic latitudes, resulting in decreasing overall completeness of
the 2MASS catalogue (e.g. Jarrett \etal 2000$b$) and consequently the
2MRS sample. 
Stellar confusion also produces colour bias in the 2MASS
galaxy photometry (Cambresy, Jarrett \& Beichman 2005) but this
bias should not be significant for 2MRS due to its relatively bright 
magnitude limit.

\subsection{Magnitude and F\mbox{}l\mbox{}ux Conversions}

2MRS uses the 2MASS magnitude $K_{20}$, 
which is defined\footnote{Column 17
(k$\_$m$\_$k20fc) in the 2MASS in the Extended Source Catalogue (XSC)}
as the magnitude
inside the circular isophote corresponding to a surface brightness of
$\mu_{K_s}=20 {\rm mag}$ arcsec$^{-2}$ (e.g. Jarrett \etal 2000$a$).
The isophotal magnitudes underestimate the total luminosity by $10\%$
for the early-type and $20\%$ for the late-type galaxies (Jarrett
\etal 2003).  Following Kochanek \etal (2001, Appendix), an offset of
$\Delta=-0.20\pm0.04$ is added to the $K_{20}$ magnitudes.
The galaxy magnitudes are corrected for Galactic extinction using the
dust maps of Schlegel, Finkbeiner \& Davis (1998) and an extinction
correction coefficient of $R_K=0.35$ (Cardelli, Clayton \& Mathis 
1989). As expected, the extinction corrections are small for the 2MRS
sample. The $K_{\rm s}$ band $k$-correction is derived by Kochanek
\etal (2001) based on the stellar population models of Worthey (1994).
The k-correction of $k(z)=-6.0\log(1+z)$, is independent of galaxy
type and valid for $z\lesssim 0.25$.

The f\mbox{}l\mbox{}ux, $S$, for each galaxy is computed from the apparent magnitudes using
\begin{equation}
S=S(0\,{\rm mag})10^{-0.4(K_{20}+ZPO)}\;,
\end{equation}
where the zero point offset is $ZPO = 0.017\pm0.005$ and $S(0 {\rm
mag})=1.122\times10^{-14}\pm1.891\times10^{-16} {\rm W cm}^{-2}$ for
the $K_{\rm s}$ band (Cohen, Wheaton \& Megeath 2003).

\subsection{The Redshift Distribution and the Selection Function}\label{sec:2mass:nz}

\begin{figure}
\psfig{figure=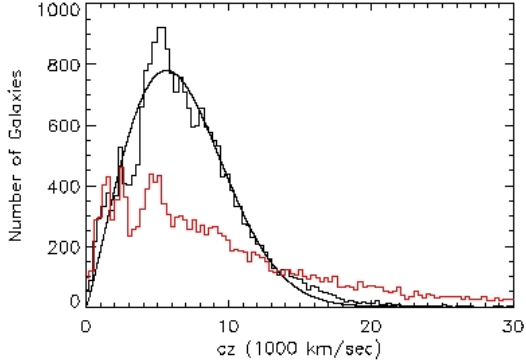,angle=0,width=0.5\textwidth,clip=}
\caption[Redshift histogram for 2MRS galaxies] {Redshift histogram for
2MRS galaxies and a least squares fit (Equation~\ref{eqn:dnz2mrs}) to
the data (black). For comparison, also plotted is a redshift histogram
for PSCz galaxies (Saunders \etal 2000) (red).}
\label{fig:2MRSnz}
\end{figure}

The redshift distribution of 2MRS is shown in
Figure~\ref{fig:2MRSnz}.  The $IRAS$ PSCz survey redshift distribution
(Saunders \etal 2000) is also plotted for comparison. 2MRS samples
the galaxy distribution better than the PSCz survey out to
$cz=15000$\kmps. The selection function of the survey (i.e. the
probability of detecting a galaxy as a function of distance) is modelled using a
parametrised fit to the redshift distribution:
\begin{equation}
dN(z)=Az^\gamma\exp\left[-\left(\frac{z}{z_c}\right)^\alpha\right]dz\;,
\label{eqn:dnz2mrs}
\end{equation}
with best-fit parameters of $A=116000\pm4000$, $\alpha=2.108\pm0.003$,
$\gamma=1.125\pm0.025$ and $z_c=0.025\pm0.001$.  This best-fit is also
shown in Figure~\ref{fig:2MRSnz} (solid line). The overall selection
function $\phi_{\rm s}(r)$ is the redshift distribution divided by the volume
element 
\begin{equation}
\phi_{\rm s}(r)=\frac{1}{\Omega_s r^2} \left(\frac{dN}{dz}\right)_r
\left(\frac{dz}{dr}\right)_r\;
\end{equation}
where $\Omega_s(\approx 4\pi$ steradians) is the solid angle of the
survey and $r$ is the comoving distance.

\section{Expansion of the Density, Velocity and Gravitational Potential Fields in Spherical Harmonics}\label{sec:2mass:sph}
This section describes the spherical harmonic expansion of the density
and velocity fields. Spherical harmonics are commonly employed in
cosmography of whole-sky surveys for two main reasons. Firstly,
separating the observed density field into angular and radial modes
concentrates the linear scale redshift distortions into one dimension, making
distortions easier to deconvolve. Secondly, both spherical
harmonic and Bessel functions are orthogonal and together they form
eigenfunctions of the Laplacian operator. 
These properties lead to very simple relationships between
the density, the velocity and the potential fields in harmonic space,
allowing the straightforward reconstruction of one field from another
observed field. The method used in this paper is based on that developed by FLHLZ. 
Therefore, we do not give a full description of our method, we 
only outline the formalism.

\subsection{Expansion of the density field}
The density field, $\rho({\bf r})$, can be expanded as a
Fourier-Bessel series in the following way:
\begin{eqnarray}
\rho({\bf r}) &=& \sum\limits_{l=0}^{l_{\rm max}}
\sum\limits_{m=-l}^{+l} \sum\limits_{n=1}^{n_{\rm max}(l)}\,  C_{ln}\,
\rho_{lmn}\, j_l(k_n r)\, Y_{lm}(\theta,\phi) \qquad \nonumber \\
&=&  \sum\limits_{lmn}\, C_{ln}\,
\rho_{lmn}\, j_l(k_n r)\, Y_{lm}(\theta,\phi). \qquad
\label{eq:denexp}
\end{eqnarray}
Here, $\rho_{lmn}$ is the density coefficient,
$Y_{lm}(\theta,\phi)$ is the spherical harmonic corresponding to the
spherical angular coordinates, $j_l(x)$ is the spherical Bessel
function and $C_{ln}$ is the normalisation constant of the spherical
Bessel Function. The
resolution of the angular modes is determined by the value of $l_{\rm
max}$ and that of the radial modes by the value of $n_{\rm max}(l)$. 
In theory, $l_{\rm max}$ and $n_{\rm max}(l)$ tend
to infinity. In practice, since the distances which 
galaxy surveys probe are finite, $l_{\rm max}$
and $n_{\rm max}(l)$ need to be truncated at finite numbers. 
This truncation is an effective smoothing of the density field and limits the 
resolution of the structure in the reconstruction such that the structures smaller
than the effective resolution of the highest harmonic,
$\Delta\theta\sim\pi/l_{\rm max}$ (cf. Peebles 1980, Section 46) and
the highest radial mode, $\Delta r\sim\pi/k_{n_{\rm max}}$, will be
lost. On the other hand, if the expansion is continued to a very high order the
resolution of expansion will be higher than the size of the real
structure, introducing artifacts into the reconstruction of density
field.  The choice of $l_{\rm max}$ and $n_{\rm max}(l)$ is discussed
in Section 6.

The radial modes, $k_n$, correspond to a given angular mode $l$ and the $C_{ln}$s 
are chosen according to the
boundary conditions and to ensure the orthogonality of the spherical
Bessel functions. It is assumed
that the data are only given within a sphere of radius $R$ and that 
the logarithmic derivative of the gravitational potential is continuous at $r=R$. 
The corresponding value of $C_{ln}$ is
$1/\left(\frac{R^3}{2}[j_{l}(k_nR)]^2\right)$.  The derivation of this
result is given in FLHLZ, Appendix A.

Given that the data from a f\mbox{}l\mbox{}ux-limited redshift catalogue are a set
of N discrete points with a redshift-space position vector {\bf s} and
a weighting function ${w(s_i)}$, the harmonic density coefficients
in redshift space can be derived using the inversion formula,
\begin{equation}
\rho_{lmn}^\ss = \sum\limits_{i=1}^N w(s_i) j_l(k_ns_i)
Y^\ast_{lm}(\theta_i, \phi_i ). \qquad
\end{equation}
where the superscripts $S$ and $R$ hereafter denote redshift and
real-space, respectively. We give each galaxy the weight
$w(s_i)=1/{\phi_{\rm s}(s_i)}$ and ${\phi_{\rm s}(s_i)}(\approx{\phi_{\rm s}(r_i)})$ is the
selection function of galaxy $i$.  

The density fluctuations,
$\delta({\bf s}) = (\rho({\bf s}) / {\bar\rho}) - 1$, can be estimated 
in terms of the Fourier-Bessel function 
\begin{equation}
\delta_{lmn}^\ss = \frac{\rho_{lmn}^\ss}{\bar\rho}-{\bf M}_{lmn},
\label{eq:deltalmn}
\end{equation}
where ${\bar\rho}$ is the mean density of the survey and the second
term ${\bf M}_{lmn}$ represents a monopole correction (cf. FLHLZ,
Equation D14).

\subsection{Expansion of the velocity field}
The linear theory velocity field in spherical harmonics is given by
Reg\"os \& Szalay (1989). It is derived by expanding the gravitational
potential $\phi({\bf r})$ and the density fluctuations $\delta({\bf
r})$ in Fourier-Bessel series, and substituting these into Poisson's
equation. Since
$j_l(k_nr)Y_{lm}(\theta,\phi)$ is an eigenfunction of the Laplacian
operator,
\begin{equation}
\nabla^2[j_l(k_nr)Y_{lm}(\theta,\phi)]=-k_n^2
j_l(k_nr)Y_{lm}(\theta,\phi),
\end{equation}
the harmonics of the gravity field can be related to those of
the galaxy density field as:
\begin{equation}
\phi_{lmn}=-\frac{3}{2} \Omega_{\rm m}^{0.6}H_0^2
\frac{\delta_{lmn}}{k_n^2}. \qquad
\end{equation}

The radial component of the
velocity field, ${\bf v}_\parallel({\bf r})=\hat{\bf r}\cdot{\bf v}({\bf r})$, 
can be decomposed into harmonics as in Equation~\ref{eq:denexp}. 
The harmonic coefficients $v_{lmn}$ for the predicted radial velocity field within
$r<R$ are derived by comparing the harmonic expansions of the velocity field and the gravitational potential field:
\begin{equation}
v_{lmn} =  \beta H_0
\sum\limits_\np \,\left( {\bf \Xi}_l\right)_{nn^\prime}
(\delta_{lmn^\prime}^\rr),
\label{eq:vlmn}
\end{equation}
with ${\bf \Xi}$ as the velocity matrix 
(cf. FLHLZ, Equation 22).

In spherical harmonics, the LG velocity,
${\bf v}({\bf 0})$, is written as (see FLHLZ, Appendix C1):
\begin{eqnarray}
{\bf v}({\bf 0}) &=& {{\beta
H_0}\over{3\sqrt{4\pi}}}\, \sum\limits_{n=1}^{n_{\rm
max}(l)}\,{C_{1n}}\, \int\limits_{0}^{R} dr^\prime j_l(k_nr^\prime)\nonumber\\ & \times &
\left(-\sqrt{2} Re[\delta^\rr_{11n}]\,{\bf i} +\sqrt{2}
Im[\delta^\rr_{11n}]\, {\bf j} + Re[\delta^\rr_{10n}]\, {\bf k}\right)
\label{eq:v(r)LGsph}
\end{eqnarray}
where $\Re[a]$ and $\Im[a]$ refer to the real and imaginary parts of a
complex number, $a$, and $R$ is the radius out to which the density
field is measured for the calculation of the velocity. 

The harmonics of the transverse velocity field (${\bf v}_\perp=-{\bf
r}\times\left[{\bf r}\times{\bf v}({\bf r})\right]$) are also related
to those of the density field.  They can be derived using quantum
mechanical methods (see FLHLZ, Appendix C2).

\section{Deconvolving Redshift-Space Distortions in Spherical Harmonics}\label{sec:2mass:zdistort}
The galaxy peculiar velocities will only introduce distortions in the radial
direction, leaving the angular parts intact. Thus, if the density field in
redshift-space is expressed in terms of the orthogonal radial and the angular
components, as it is the case in the Fourier-Bessel space, the redshift
distortions will only couple the radial modes of the real space
density field. This coupling is described by a coupling matrix,
$\left({\bf Z}_l\right)_{nn^\prime}$ (Fisher \etal 1994 and FLHLZ):
\begin{equation}
\delta_{lmn}^\ss = \sum\limits_{n^{\prime}=1}^{n_{\rm max}(l)}\, ({\bf
Z}_l)_{nn^{\prime}}\, \delta_{lm{n^\prime}}^\rr, \qquad
\label{eq:rhostorhor}
\end{equation}
where $({\bf Z}_l)_{nn^{\prime}}$ depends only on $H_0$,
$\beta$, $w(r)$ and $\phi_{\rm s}(r)$ and is independent of the power spectrum\footnote{The coupling matrix has a more
complicated dependence on $l$ and $m$ for catalogues with incomplete
sky coverage.  Zaroubi \& Hoffman (1996) derived a similar expression
for the coupling matrix in Cartesian coordinates.}.
The full derivation of the coupling matrix is given in FLHLZ, Appendix
D. 
We note that Equation~\ref{eq:rhostorhor} is calculated for linear
scales only. On small scales, the peculiar velocities introduce radially extended 
distortions called {\it Fingers of
God}. These distortions can be corrected for either by collapsing the fingers to a 
point at high density regions or by assuming a Maxwellian distribution for the peculiar velocities (Peacock \& Dodds 1994) 
 and deconvolving this distribution from the reconstructed fields (Heavens \& Taylor 1995; Erdo\u{g}du \etal 2004).
In any case, they will not change the constructed fields
substantially as they are smoothed out
due to the resolution of the spherical harmonics (see Erdo\u{g}du \etal 2004 who use who a similar resolution).

In the absence of shot-noise, the real-space density harmonics can
simply be obtained by inverting the coupling matrix in
Equation~\ref{eq:rhostorhor}.  However, in the presence of shot-noise,
a straight inversion becomes unstable leading to inaccurate estimation
of the real-space harmonics. This problem can be overcome by Wiener
filtering.

\section{Wiener Reconstruction}\label{sec:2mass:wf} 
In this section, we give the formu\mbox{}l\mbox{}a\mbox{}e for the Wiener filter method. For a full derivation, we refer the reader to Zaroubi \etal (1995) and Erdo\u{g}du \etal (2004).

\subsection{Model Signal and Noise Matrices}
\label{subsec:2mass:nsig}
For this analysis, the Wiener-filter reconstructed real-space
density field is given by:
\begin{equation}
{(\delta_{lmn}^\rr)}_{\rm{WF}} = \sum\limits_{\np\npp} \left({\bf S}_l
\left[ {\bf S}_l + {\bf N}_l \right]^{-1}\right)_{n\np} \left({\bf
Z}_l^{-1}\right)_{\np\npp} \delta_{lm\npp}^\ss \qquad.
\label{eq:rholmnwf}
\end{equation}
For the case $w(r)=1/\phi_{\rm s}(r)$, the signal and noise matrices simplify to
\begin{equation}
({\bf S}_{l})_{n\np} \simeq P(k_n)C_{ln}^{-1}\delta_{nn^\prime}^K\; and
\label{eq:S} 
\end{equation}
\begin{equation}
({\bf N}_l)_{n\np}=
\frac{1}{\bar\rho}\int\limits_{0}^{R} dr r^2 \frac{1}{\phi_{\rm s}(r)} 
j_{l}(k_{n}r)j_{l}(k_{\np}r)\;,
\label{eq:Nlmn}
\end{equation}
respectively. 

The expected scatter in the reconstructed density, $\langle\Delta\delta({\bf r})\rangle^2$ is given by
\begin{eqnarray}
\langle \Delta\delta({\bf r}) \rangle^2&=& \frac{1}{4\pi}\sum\limits_{ln\np} (2l+1) [({\bf I}-{\bf F}_l){\bf S}_l]_{n\np}\nonumber \\ 
&\times&C_{ln}C_{l\np}j_l(k_nr)j_l(k_\np r)\;.
\label{eq:dscatter}
\end{eqnarray}
The scatter in the reconstructed radial 
velocity $\langle\Delta {\bf v}_r({\bf r})\rangle^2$ fields can also be
formulated in Fourier-Bessel space in an analogous way 
(cf. FLHLZ, Appendix F):  
\begin{eqnarray}
\langle \Delta {\bf v}_r({\bf r}) \rangle^2&=& \frac{H_0\beta^2}{4\pi}
\sum\limits_{ln\np} (2l+1) [({\bf I}-{\bf F}_l){\bf S}_l]_{n\np}\nonumber \\ 
&\times&C_{ln}C_{l\np}
\frac{j^\prime_l(k_nr)j^\prime_l(k_\np r)}{k_n k_\np}.
\label{eq:vscatter}
\end{eqnarray}
\subsection{Choice of Reference Frame}
The redshift distortion matrix given in Section 4 
takes into account the motion of the observer, ${\bf v(\bf 0)}$, and 
thus the choice of reference frame is arbitrary. 
On the other hand, the formulation of $\left({\bf Z}_l\right)_{nn^\prime}$  
involves the Taylor expansion of the velocity field out to first order in 
$\Delta{\bf v}={\bf v(\bf r)}-{\bf v(\bf 0)}$ (see Appendix D of FLHLZ). 
As such, the reconstruction will be more accurate in the frame in which 
$\Delta{\bf v}$ is smaller. Nearby where the galaxies 
share the LG motion, it is more accurate to  
correct for the redshift distortions in the LG frame. 
At larger distances, galaxy motions are independent of the LG velocity and thus  
$\Delta{\bf v}$ is smaller in the CMB frame. Since 2MRS 
probes the local universe 
($cz_{\rm med}\approx6000$\kmps), we will work with the LG frame redshifts. 
We will present the dipole velocity results from redshifts in both frames. 

The choice of frame for the redshifts will not affect the frame of the 
reconstructed density field. In both cases, the reconstructed harmonics 
will be in real-space. Consequently, the 
reconstructed velocity field (Equation~\ref{eq:vlmn}) will be in the CMB frame. 
It is possible to convert the velocity field into the LG frame by subtracting the Local Group velocity ($v_{LG}=627\pm22$
\kmps, towards $l_{LG}=273^\circ\pm3^\circ$,
$b_{LG}=29^\circ\pm3^\circ$, 
Bennett \etal 2003 and Courteau \& Van Den Bergh
1999). 
In the following section, 
the reconstructed 
velocity maps will be presented in either the LG or the CMB frames. 
Up to 8000 \kmps (Figures~\ref{fig:velshells(a)},~\ref{fig:velshells(b)},~\ref{fig:velshells(c)} and ~\ref{fig:velshells(d)})      
it is difficult to see the real nature of the structures in the LG frame as the LG velocity dominates. Therefore, the maps will be plotted in the CMB frame.
On larger distances (Figures~\ref{fig:velshells(e)},~\ref{fig:velshells(f)},~\ref{fig:velshells(g)} and ~\ref{fig:velshells(h)}), galaxies have positive CMB radial 
velocities so it is easier to compare the velocities in the LG frame than in the CMB frame.

\section{Application to the Two Mass Redshift Survey}\label{sec:2mass:app} 
The formalism discussed in the previous sections is limited to $4\pi$
sky coverage. In principle, the analysis can be extended to account
explicitly for incomplete sky coverage. In this paper, a simpler
approach has been adopted to fill in the galaxies masked by the Galactic 
Plane (the {\it Zone of Avoidance}). The masked 
area is divided into 36 bins of $10^\circ$ in
longitude.  In each angular bin, the distance is divided into bins of 10
$h^{-1}$ Mpc. The galaxies in each longitude/distance bin are then
sampled from the corresponding longitude/distance bins in the adjacent
strips $-|b_{\rm masked}|-10^\circ<b<|b_{\rm masked}|+10^\circ$ (where
$|b_{\rm masked}|=5^\circ$ or $|b_{\rm masked}|=10^\circ$).  These galaxies
are then placed in random latitudes within the mask region. The
number of galaxies in each masked bin is set to a random Poisson
deviate whose mean equals the mean number of galaxies in the
adjacent unmasked strips. This procedure is carried out to mimic the
shot noise effects. The success of this interpolation method depends on the 
interplay between the width of the mask, the angular resolution and whether the structure in the adjacent regions physically correlate with the structure in the Zone of Avoidance. 
The method is less robust for masked areas larger than $|b|=15^\circ$
(Lahav \etal 1994). 
The width of the 2MRS mask is much smaller than this value and 
the survey penetrates deep into the Zone
of Avoidance apart from very obscured regions near the centre of the
Milky Way.  We also test the effects of our method by varying  
the bin sizes and by filling the region uniformly instead of interpolating.
We calculate the angular power-spectrum for each case and find good agreement 
(Rassat \etal 2006) between them. 

The density field is expanded within a spherical volume of radius
$R_{\rm max}=20000 \kmps$. We impose the condition that the density field
outside $R_{\rm max}$ is zero and the logarithmic 
derivative of the gravitational potential is continuous at the boundary 
(which holds if $j_{\rm l-1}(k_nR)=0$ for all $l$). 
There are two other possible 
boundary conditions discussed in FLHLZ: setting only the density or the 
radial velocity zero at the boundary. 
We have tested our reconstructions using these boundary conditions.
Setting the density to
zero at the boundary results in a discontinuous potential and 
the reconstructed radial velocities become
unrealistically high. Up to a
distance of  
10000 $\kmps$, the fields reconstructed using the 
zero velocity boundary condition agree reasonably well with those
reconstructed using the continuous potential condition, however
the radial velocities of the zero velocity reconstructions are
slightly higher.
Beyond 10000 $\kmps$, the radial velocities of the 
zero velocity reconstructions tend to zero, however the backside infall into
the Shapley Supercluster still exists. 
Thus we conclude that setting the logarithmic 
derivative of the gravitational potential to be continuous at the boundary
allows for a {\it smooth} transition of the density and velocity fields
from $r < R_{\rm max}$ to $r > R_{\rm max}$ and as such gives the most robust results. We also note that FLHLZ, WLF and Schmoldt \etal (1999) use the same boundary condition for their reconstructions. 

The resolution of the 
reconstruction depends on the maximum values of angular and radial modes (see 
Figure 6 of WLF). 
We want to have as high a resolution as possible without introducing 
artificial structures to the field.
Therefore, we compute the harmonics up to $l_{\rm
max}=15$ angularly and $k_nR\leq100$ radially. $l_{\rm max}=15$ is
chosen so that the angular resolution ($\delta\theta\sim\pi/l$) is
comparable to the width of the interpolated region and the radial modes
are chosen to match the angular resolution 
(see FLHLZ, Appendix B for a detailed discussion).

The Wiener filter method requires a model for the linear galaxy power
spectrum in redshift-space, which depends on the real-space power
spectrum and on the redshift distortion parameter,
$\beta\equiv\omegam^{0.6}/b$.  
The real-space power spectrum is
well described by a scale-invariant Cold Dark Matter power spectrum
with shape parameter $\Gamma\approx\omegam h$, 
for the scales concerned in this
analysis. The combined analysis of
third year data from the Wilkinson Microwave Anisotropy Probe (WMAP, Spergel \etal 2006), the Two Degree Field Galaxy Redshift Survey 
(2dFGRS, Cole \etal 2005) and the 
Sloan Digital Sky Survey (Tegmark \etal 2004)  
imply $\Gamma\approx0.17$. On the other hand 
Frith, Outram \& Shanks (2005) and Maller \etal (2005) analyse 
the angular power spectrum of the 2MASS galaxies and 
derive lower $\Gamma$ values of $0.14\pm0.02$ and $0.116\pm0.009$, respectively.
The normalisation of the
power spectrum is conventionally expressed in terms of the variance of
the mass density field in spheres of $8 h^{-1} $ Mpc, $\sigma_{8}$. Recent 
WMAP results combined with the 2dFGRS gives $\sigma_{8}\approx0.75$, whereas 
analysis of the 2MASS galaxies suggest $\sigma_{8}\approx0.9$ (Frith, Outram \& Shanks 2005, Maller \etal 2005, Pike \& Hudson 2005).
In a previous paper, we analysed the 2MRS LG dipole and derived a 
value for $\beta$ for 2MRS: $\beta=0.4\pm0.09$ (Erdo\u{g}du \etal 2006). 

In order to check the
goodness of fit of these different parameters to the data, 
we calculate the $\chi^2$
statistic defined by
\begin{equation}
\chi^2={\bf d}^\dagger ({\bf S}+{\bf N})^{-1} {\bf d},
\end{equation}
for $\Gamma=(0.1,0.2)$, $\sigma_8=(0.7,0.9)$ and $\beta=(0.3,0.4,0.5)$.
Changing $\Gamma$ does not affect the $\chi^2$ values greatly but $\Gamma=0.2$ 
is a slightly better fit. Varying $\sigma_8$ is a bit more influential on the $\chi^2$ values, however, the reconstructed maps look very similar for these models.
Increasing $\Gamma$ decreases the
power on large scales and, therefore, the peculiar velocities tend to be
smaller but the difference is negligible.
Decreasing $\sigma_{8}$ has the same effect on the power
spectrum as increasing $\Gamma$.  
The choice of $\beta$ plays a more important role in the goodness-of-fit results but has little affect on the reconstructed maps other than a linear scaling of peculiar velocities. 
In the analysis that follows, we adopt $\Gamma=0.2$, $\sigma_8=0.7$ and $\beta=0.5$, which is the best fit model that was considered, with a 
reduced $\chi^2$ value of 1.3. We note that the best fit $\beta$ is higher than derived in Erdo\u{g}du \etal (2006), this discrepancy will be discussed 
in Section 6.3. 
In a forthcoming paper (Rassat \etal 2006), we will present a detailed analysis of the cosmological parameters of 2MRS. 

\subsection{Density Maps}
\begin{figure*}
\psfig{figure=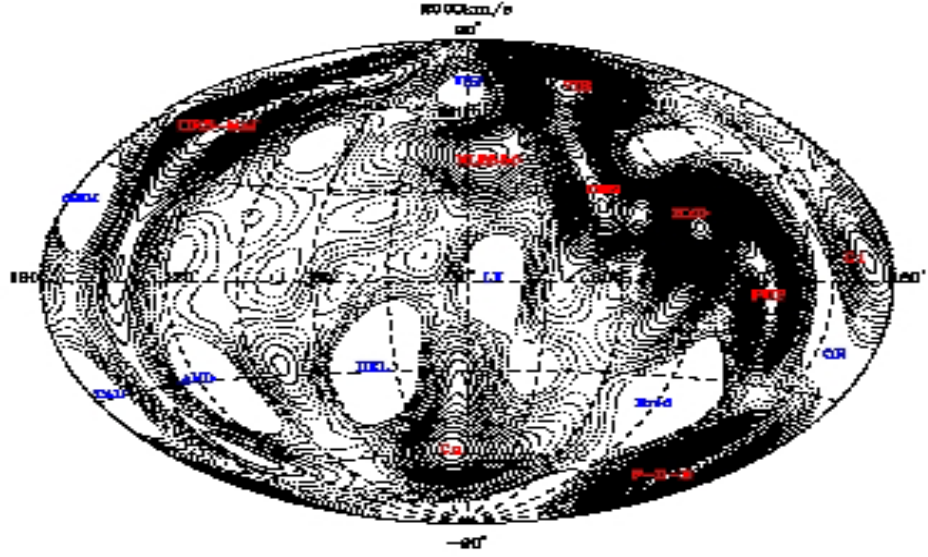,angle=90,width=\textwidth,height=0.4\textheight,clip=}
\caption[The reconstructed density fields, evaluated on a thin shell at
 $2000 \kmps$] {The reconstructed density field, evaluated on a thin
 shell at $2000 \kmps$, shown in Galactic Aitoff projection.  Dashed
 lines show $\delta < 0$, and solid lines show $\delta \ge 0$, with
 contour spacing $|\Delta \delta| = 0.1$. The overdense regions are 
(from left to right)
Ursa Major (Urs-Maj),WLB 550, ${\rm C}\alpha$, Virgo (Vir), 
Centaurus (Cen), Hydra (Hyd), Puppis (Pup),
 Fornax-Doradus-Eridanus (F-D-E) and C1 .  The voids are Gemini (Gem), 
Taurus (Tau), Andromeda (And), Delphinus (Del), Virgo (Vir), Local Void (LV), 
Eridanus (Erid) and Orion (Or).}
\label{fig:denshells(a)}
\end{figure*}

\begin{figure*}
\psfig{figure=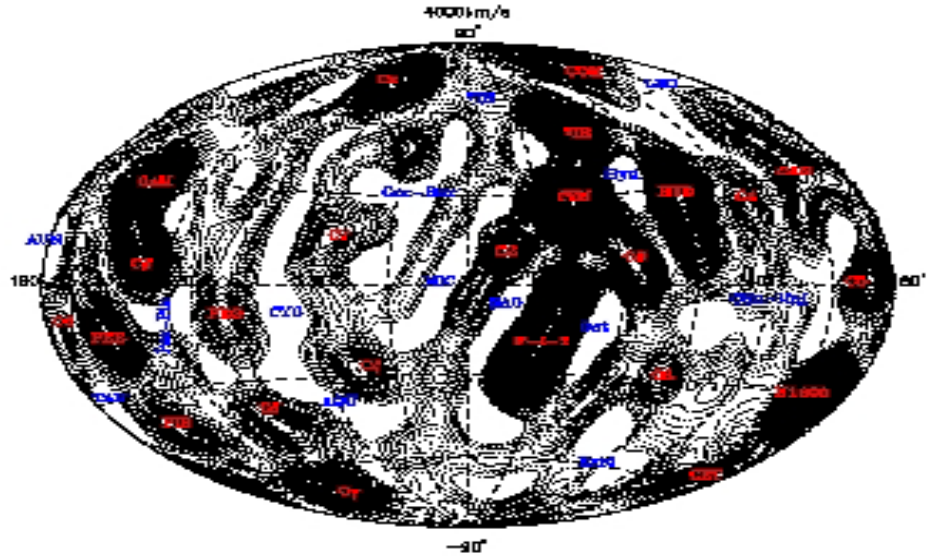,angle=90,width=\textwidth,height=0.4\textheight,clip=}
\caption[] { 
Same as Figure~\ref{fig:denshells(a)} but evaluated
 at $4000 \kmps$. The overdense regions are  C6, Perseus (Per), Pisces (Pis),
Camelopardalis (Cam), C$\beta$, Pegasus (Peg), C$\delta$, C$\gamma$, 
C$\epsilon$, C7, C$\zeta$, C2, Coma (Com), Virgo (Vir), 
Centaurus (Cen), Pavo-Indus-Telescopium (P-I-T), C8, Hydra (Hyd), 
C3, C4, Cancer (Can), NGC 1600 (N1600), Cetus (Cet) and C5. The voids are Aunga (Aun), Taurus (Tau), Perseus-Pisces (Per-Pis), Cygnus (Cyg), Aquarius (Aqu), Corona Borealis (Cor-Bor), Microscopium (Mic), Virgo (Vir), Sagittarius (Sag), Hydra (Hyd), Octans (Oct), Leo, Eridanus (Erid) and Canis Major (Can-Maj).}
\label{fig:denshells(b)}
\end{figure*}

\begin{figure*}
\psfig{figure=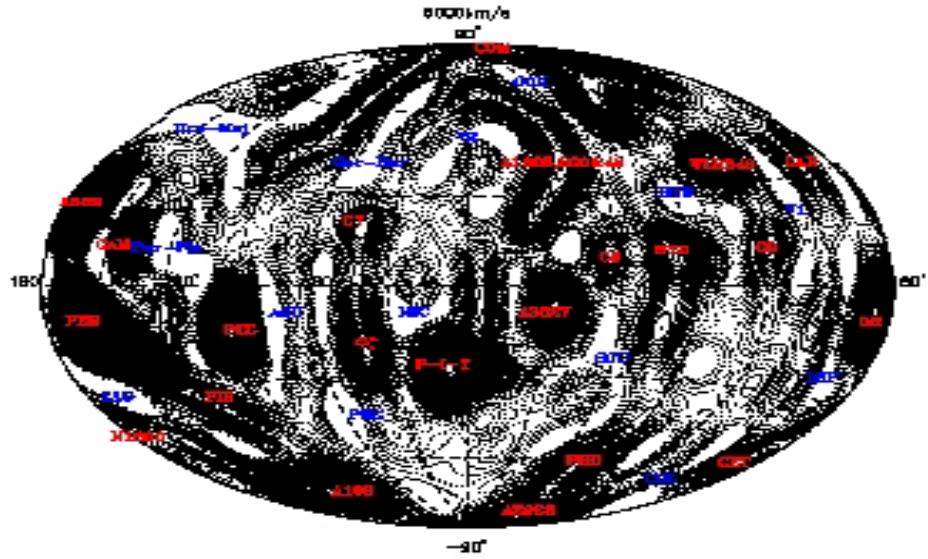,angle=90,width=\textwidth,height=0.4\textheight,clip=}
\caption[] {Same as in Figure~\ref{fig:denshells(a)} but evaluated
 at $6000 \kmps$. The overdensities are Abell 569 (A569), Perseus (Per), Pisces (Pis), NGC 1600 (N1600), Abell 168, Camelopardalis (Cam),  Pegasus (Peg), C7, C$\zeta$,  Pavo-Indus-Telescopium (P-I-T), Abell Clusters 1605, 300 \& 42,  Abell 3627, C8, Phoenix (Pho), Abell 2923, Coma (Com), Hydra (Hyd), 
WLB 248, C9, Cancer (Can), Orion (Or) and Cetus (Cet). The voids
 are Taurus (Tau), Ursa Major (Urs-Maj), Perseus-Pisces (Per-Pis), Aquarius (Aqu), Corona-Borealis (Cor-Bor), Pegasus (Peg), V2, Microscopium (Mic), 
Coma (Com),  Sculptor (Scu), Columba (Col), V1 and Lepus (Lep).}
\label{fig:denshells(c)}
\end{figure*}

\begin{figure*}
\psfig{figure=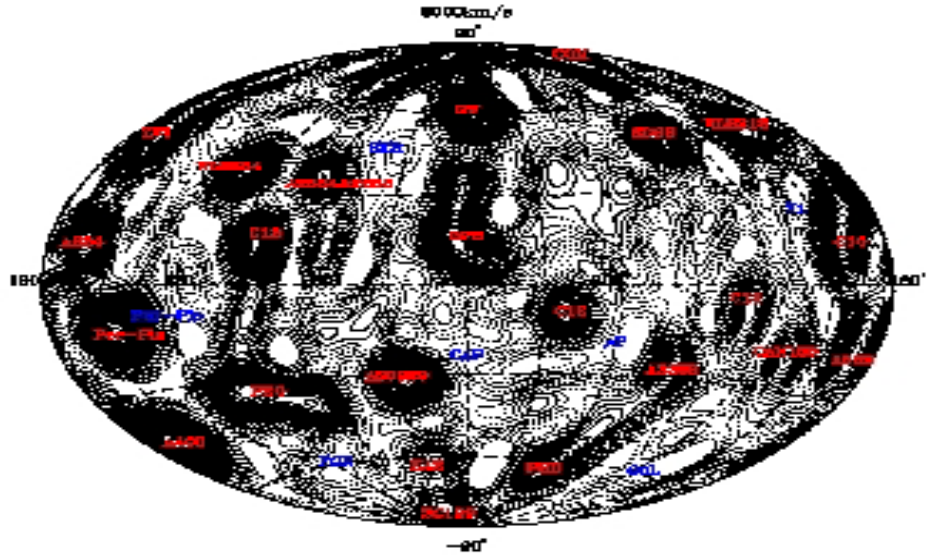,angle=90,width=\textwidth,height=0.4\textheight,clip=}
\caption[] {Same as in Figure~\ref{fig:denshells(a)} but evaluated
   at $8000 \kmps$. The overdensities are ZwCl 0943.7+5454 (ZWI), 
Abell 634, Perseus-Pisces (Per-Pis), Abell 400, WLB 664, C13, Pegasus (Peg), 
Abell Clusters 2634 \& 2666, Abell S0929, Klemola 42 (K42), 
Great Wall (GW), Ophiuchus (Oph), SC 129, C12, Phoenix (Pho), Coma (Com), SDSS CE J159.778641-00.784376 (SDSS), Abell 3389, WLB 213, C11, CAN 136, C10 and Abell 539. 
The voids are Perseus-Pisces (Per-Pis), Fornax (For), Serpens (Ser), Capricornus (Cap), Apus (Ap), Columba (Col) and V1.}
\label{fig:denshells(d)}
\end{figure*}

\begin{figure*}
\psfig{figure=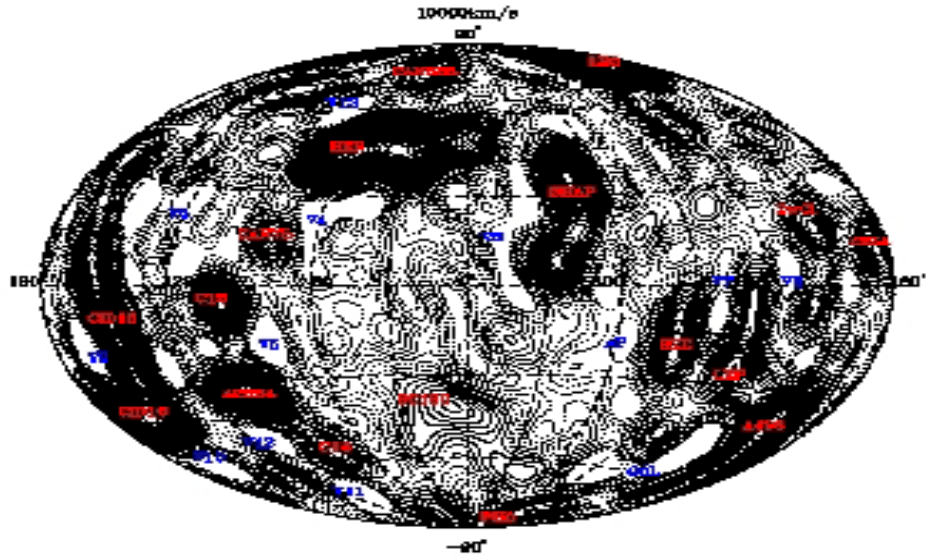,angle=90,width=\textwidth,height=0.4\textheight,clip=}
\caption[] {Same as in Figure~\ref{fig:denshells(a)} but
 evaluated at $10000 \kmps$.  The overdensities are CID 15, CID16, C13, Abell 2634, C14, CAN 75, CAN 338, Hercules (Her), SC180, Shapley (Shap), Phoenix (Pho), Leo, RXC J0712.0-6030, Lepus (Lep), ZwCl 0820.6+0436, Abell 496 and CIZA J0603.8+2939. The voids are V9, V6, V10, V12, V11, V13, V5, V4, V3, Apus (Ap), V7, Columba(Col) and V8.}
\label{fig:denshells(e)}
\end{figure*}

\begin{figure*}
\psfig{figure=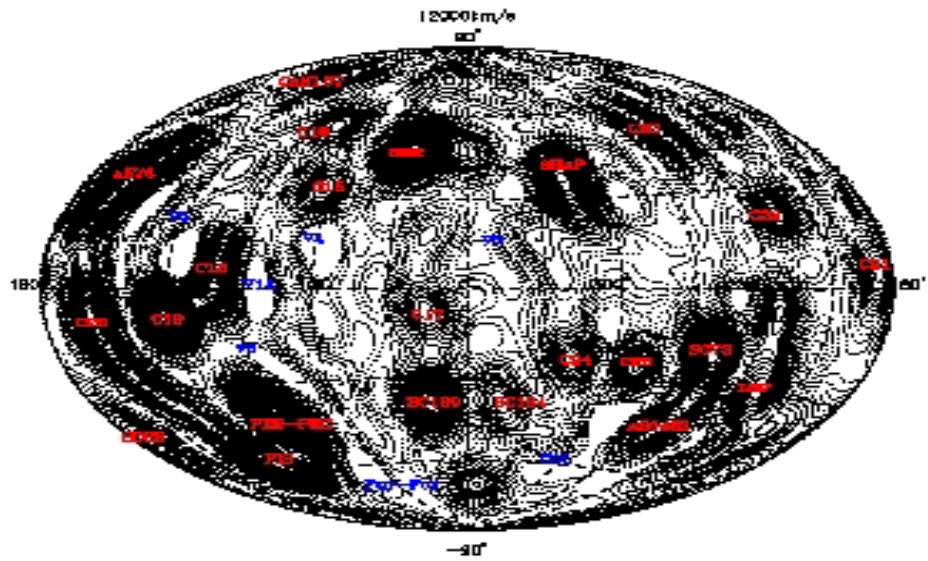,angle=90,width=\textwidth,height=0.4\textheight,clip=}
\caption[] {Same as in Figure~\ref{fig:denshells(a)} but
 evaluated at $12000 \kmps$.  The overdensities are SDSS clusters SDSS CE J048.551922-00.613762 and SDSS CE J050.935932-00.117944 (SDSS), Abell 576, C20, Pisces (Pis), CAN 137, C19, C18, Perseus-Pegasus (Per-Peg), C16, C15, Hercules (Her),  C17, SC180, SC194, Shapley (Shap), C24, C23, Leo, SC73, Abell S0463, C22, Lepus (Lep) and C21. 
The voids are V6, V14, V5,  V4, Further-Fornax (Fur-For), V3 and Columba (Col).}
\label{fig:denshells(f)}
\end{figure*}

\begin{figure*}
\psfig{figure=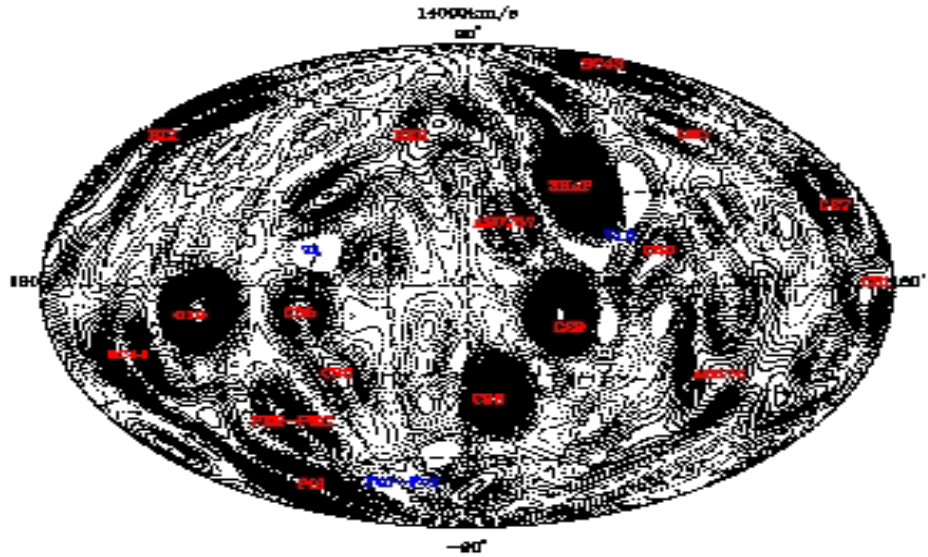,angle=90,width=\textwidth,height=0.4\textheight,clip=}
\caption[The reconstructed density fields, evaluated on a thin shell at
 $14000 \kmps$ ] {Same as in Figure~\ref{fig:denshells(a)} but
 evaluated at $14000 \kmps$. The overdensities are Rixos F231\_526 (RIX), SC44, C19, Pisces (Pis), Perseus-Pegasus (Per-Peg), C25, C26, Hercules (Her), Abell S0757, C28, Shapley (Shap), C29, C30, SC 43, Leo, Abell 3376, C27 and C21.  
The voids are V4, Further-Fornax (Fur-For) and V15.}
\label{fig:denshells(g)}
\end{figure*}

\begin{figure*}
\psfig{figure=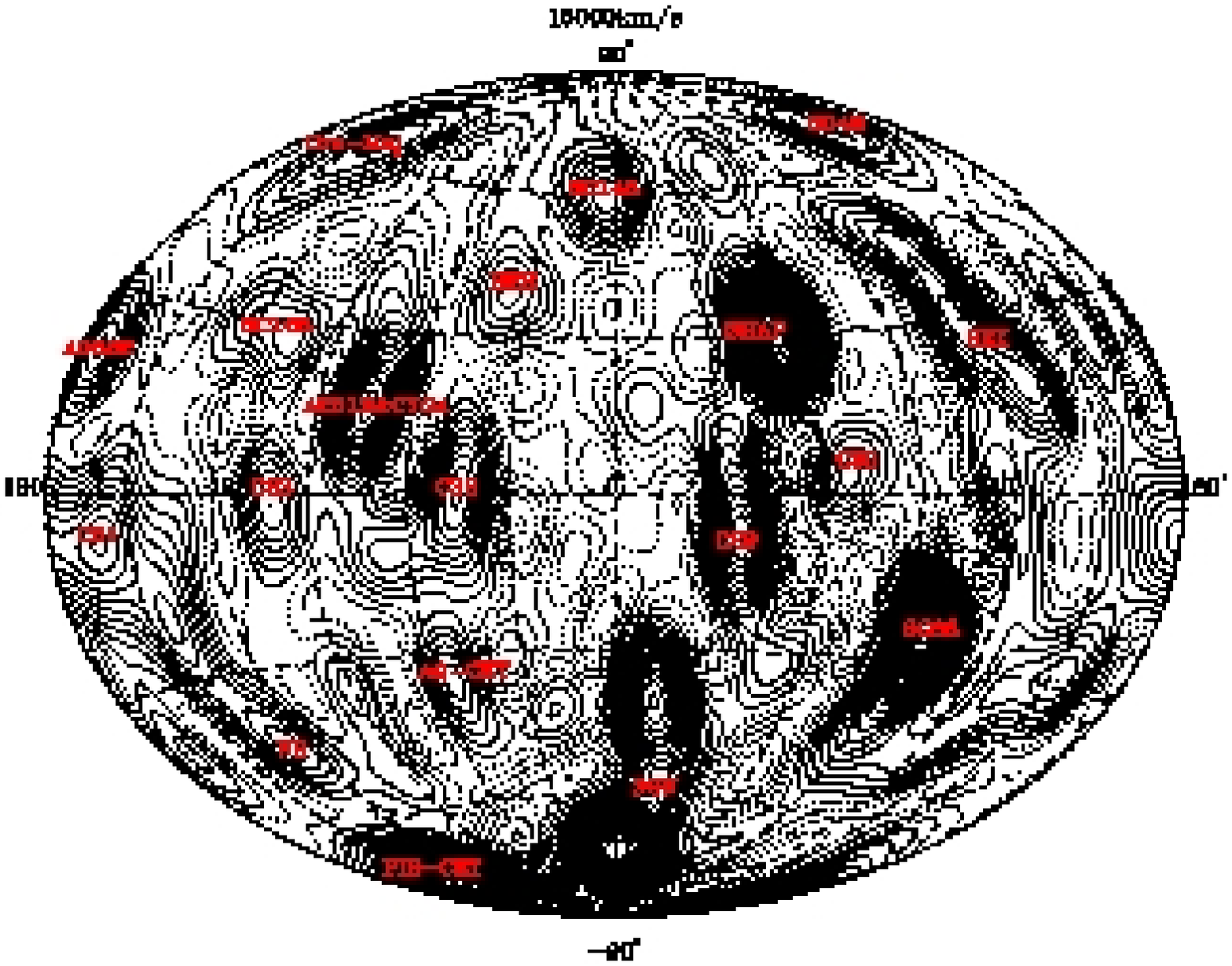,angle=90,width=\textwidth,height=0.4\textheight,clip=}
\caption[] {Same as in Figure~\ref{fig:denshells(a)} but
 evaluated at $16000 \kmps$. The overdensities are Abell 582, C31, Ursa-Major (Urs-Maj), SC168, West 9 (W9), C33, Abell 2319 \& Cygnus A (CYGA), Pisces-Cetus (Pis-Cet), C32, Aquarius-Cetus (Aq-Cet), Hercules (Her), SC143, Southern Great Wall (SGW), C29, Shapley (Shap), C30, SC43, Sextans (Sex) and SC68.}
\label{fig:denshells(h)}
\end{figure*}

Figures~\ref{fig:denshells(a)},~\ref{fig:denshells(b)},~\ref{fig:denshells(c)},~\ref{fig:denshells(d)},~\ref{fig:denshells(e)},~\ref{fig:denshells(f)},~\ref{fig:denshells(g)} and~\ref{fig:denshells(h)} show the Aitoff
projections of the reconstructed density field in real-space, plotted
in Galactic coordinates, evaluated at different
distance slices. 

{\bf Figure~\ref{fig:denshells(a)}} shows a thin density shell at
$r=2000 \kmps$. The expected
scatter in the reconstructed density field is
$\langle\Delta\delta({\bf r})\rangle=0.31$ and this value 
stays approximately the same out to $r=10000 \kmps$ 
and increases by 0.1 thereafter. 
All the known structures are resolved and the field at this distance 
is dominated by inter-connecting voids. 
The Puppis cluster (Lahav \etal 1993, S1 in Saunders \etal 1991) is very prominent and 
extends down to $b\simeq-28^\circ$. The overdensity 
labelled as C1 ($l\simeq190^\circ$ ,$b\simeq5^\circ$) does not correspond 
to any previously observed structure. 

{\bf Figure~\ref{fig:denshells(b)}} is the 2MRS density field at
$r=4000 \kmps$.   Hydra ($l\simeq270^\circ$, $b\simeq50^\circ$) and Centaurus ($l\simeq300^\circ$, $b\simeq20^\circ$) continue in this shell but now are separated by the Hydra
void. Centaurus is connected to the new overdensities C2 ($l\simeq345^\circ$, $b\simeq10^\circ$) and C8 ($l\simeq295^\circ$, $b\simeq7^\circ$) and 
Hydra connects to C4 ($l\simeq240^\circ$, $b\simeq25^\circ$). 
Pavo-Indus-Telescopium continues deep into the Zone of Avoidance and connects to the Centaurus Wall as 
suggested by Fairall (1988). The well-known cluster 
Cancer ($l\simeq210^\circ$, $b\simeq30^\circ$), as well as new
clusters C3 ($l\simeq275^\circ$, $b\simeq-30^\circ$), 
C5 ($l\simeq195^\circ$, $b\simeq0^\circ$), 
C6 ($l\simeq170^\circ$, $b\simeq-15^\circ$) and C7 ($l\simeq55^\circ$, $b\simeq15^\circ$), all peak at $r=5000 \kmps$ with  
C5, C6 and C7 connecting to Pegasus and the Perseus-Pisces Wall. The overdensity C$\epsilon$ (WLF) is more prominent in the 2MRS field than in $IRAS$ 1.2 Jy field.
The Perseus and Pisces clusters are picked up
as two distinct overdensities and both connect to 
Camelopardalis ($l\simeq145^\circ$ ,$b\simeq30^\circ$) and Cetus
($l\simeq190^\circ$, $b\simeq-55^\circ$) at $r=5000 \kmps$. 

{\bf Figure~\ref{fig:denshells(c)}} is the density field at
$r=6000 \kmps$. Hydra ($l\simeq285^\circ$, $b\simeq5^\circ$) is less
prominent in this shell whereas Pavo-Indus-Telescopium
($l\simeq0^\circ$, $b\simeq-45^\circ$) is still strong. The 
cluster Abell 3627 (or Norma, $l\simeq330^\circ$, $b\simeq-10^\circ$) 
which was recognised as the centre of the Great Attractor\footnote{By `Great Attractor', it is meant the entire
steradian on the sky centred at ($l\sim310^\circ$,$b\sim20^\circ$)
covering a distance of 20 $h^{-1}$ Mpc to 60 $h^{-1}$ Mpc.} by Kraan-Korteweg \etal (1996), peaks around $r=5000 \kmps$ and is separated from the 
Centaurus Wall by the Sculptor Void. The overdensity labelled as C8 continues from the previous shell and contains the luminous CIZA J1324.7-5736, cluster which was discovered by Ebeling, Mullis \& Tully (2002) and the Centaurus-Crux cluster (Woudt, 1998). 
The Abell clusters 569 ($l\simeq155^\circ$, $b\simeq-15^\circ$) 
and 168 ($l\simeq120^\circ$, $b\simeq70^\circ$) 
are as prominent as the Perseus-Pisces ridge and both are connected to it. 
The clusters labelled as WLB (e.g. WLB 
248 at $l\simeq240^\circ$, $b\simeq40^\circ$) 
were identified by White \etal (1999). There are two previously unidentified voids at this distance labelled as V1 and V2 and a cluster at ($l\simeq240^\circ$, $b\simeq8^\circ$) labelled C9.

{\bf Figure~\ref{fig:denshells(d)}} is the 2MRS density field at
$r=8000 \kmps$.  The Great Wall (Geller \& Huchra 1989) connects to 
the Ophiuchus supercluster discovered by Wakamatsu \etal
(1994). The massive Shapley supercluster is already
visible ($l\simeq300^\circ$, $b\simeq30^\circ$) at this distance. 
The void V1 continues to this shell. The new clusters C10 ($l\simeq205^\circ$, $b\simeq10^\circ$), C11 ($l\simeq250^\circ$, $b\simeq-10^\circ$) , C12 ($l\simeq325^\circ$, $b\simeq-10^\circ$) and C13 ($l\simeq85^\circ$, $b\simeq13^\circ$) lie on galactic latitudes. The cluster ZwCl 0943.7+5454 was first identified by Zwicky \& Herzog (1966) and K42 by Klemola (1969). All the superclusters 
labelled as SC (e.g. SC 129 in this shell) are given in Einasto \etal (1997, Table 2) and structures labelled CAN and CID 
(e.g. CAN 136 in this shell and CID 15 in the next shell) are clusters given in Wegner \etal (1999).

{\bf Figure~\ref{fig:denshells(e)}} is the 2MRS density field at
$r=10000 \kmps$. The expected scatter now rises to
$\langle\Delta\delta({\bf r})\rangle=0.36$. The mighty Hercules supercluster dominates this shell 
and Shapley begins to appear. Other dominant structures are CID (CAN) 15 \& 16. 

{\bf Figure~\ref{fig:denshells(f)}} is the 2MRS density field at
$r=12000 \kmps$ and has a scatter of 
$\langle\Delta\delta({\bf r})\rangle=0.41$. The Perseus-Pegasus and Pisces (Einasto \etal 1997) 
superclusters, C20 and Abell 576 continue the Perseus-Pisces wall and dominate the shell.  

{\bf Figure~\ref{fig:denshells(g)}} is the 2MRS density field at
$r=14000 \kmps$ with scatter $\langle\Delta\delta({\bf r})\rangle=0.46$.
The density field of the Shapley supercluster peaks at this distance and connects to C28 and C29 at $r=15000 \kmps$ to form a ridge.

{\bf Figure~\ref{fig:denshells(h)}} is the 2MRS density field at
$r=16000 \kmps$. The expected scatter is
$\langle\Delta\delta({\bf r})\rangle=0.49$. The most conspicuous structure is the Pisces-Cetus supercluster, which 
connects to the Southern Great Wall. 
The Perseus-Cetus also dominates the southern strips of the 2dF (see e.g. 
Erdo\u{g}du \etal 2004, Porter \& Raychaudhury 2005) and Sloan surveys  (e.g. Porter \& Raychaudhury 2005). 
We note that at this distance, we do not pick up any voids. This is due to the 
nature of our technique. 
The Wiener filter method will predict 
the mean field in the absence of data. 
Since the density field is constructed to have zero mean, 
the Wiener filter signal will approach to zero at 
the edges of the survey where the shot noise dominates. This leads to the 
signal being reconstructed in a non-uniform manner. As such, we do not present
maps of the density and velocity fields beyond 16000 $\kmps$ after which the
shot noise is too high and the Wiener Filtered fields tend to zero. 

\subsection{Velocity Maps}
\begin{figure*}
\psfig{figure=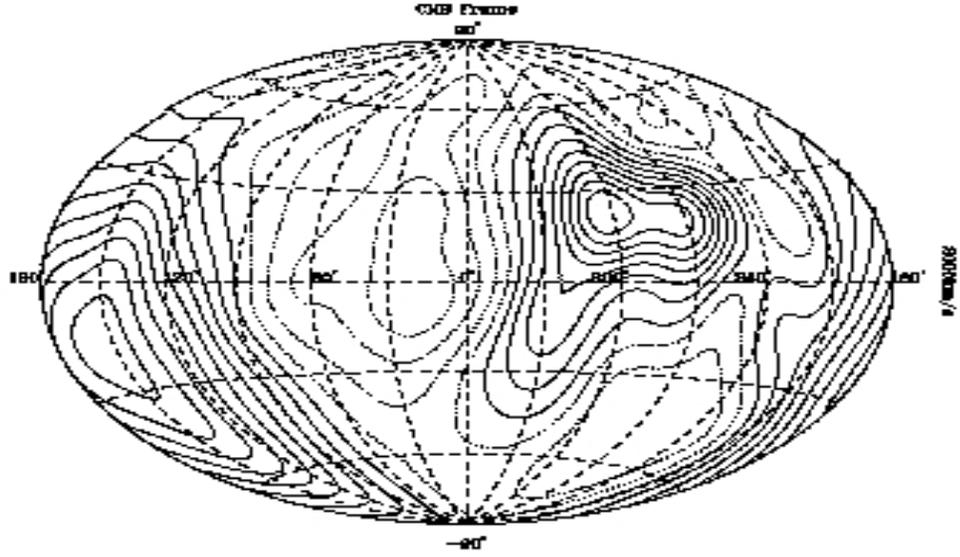,angle=90,width=\textwidth,height=0.4\textheight,clip=}
\caption[The reconstructed velocity fields, evaluated on a thin shell
 at $2000 \kmps$]{The reconstructed radial velocity field, evaluated
 on a thin shell at $2000 \kmps$, shown in Galactic Aitoff projection
 in the CMB frame.  This velocity field corresponds with the density field
 shown in Figure~\ref{fig:denshells(a)}. 
Dashed lines show infall,
 and solid lines show outflow. The first solid line is for $v_{\rm
 radial} = 0 \kmps $, and contour spacing is $|\Delta v_{\rm radial}| =
 50 \kmps$. Note the strong outflow towards the Great Attractor and
 flow out of the Local Void.  Also Perseus-Pisces and N1600, cause an
 outflow towards ($l\simeq165^\circ$, $b\simeq-30^\circ$)}.
\label{fig:velshells(a)}
\end{figure*}

\begin{figure*}
\psfig{figure=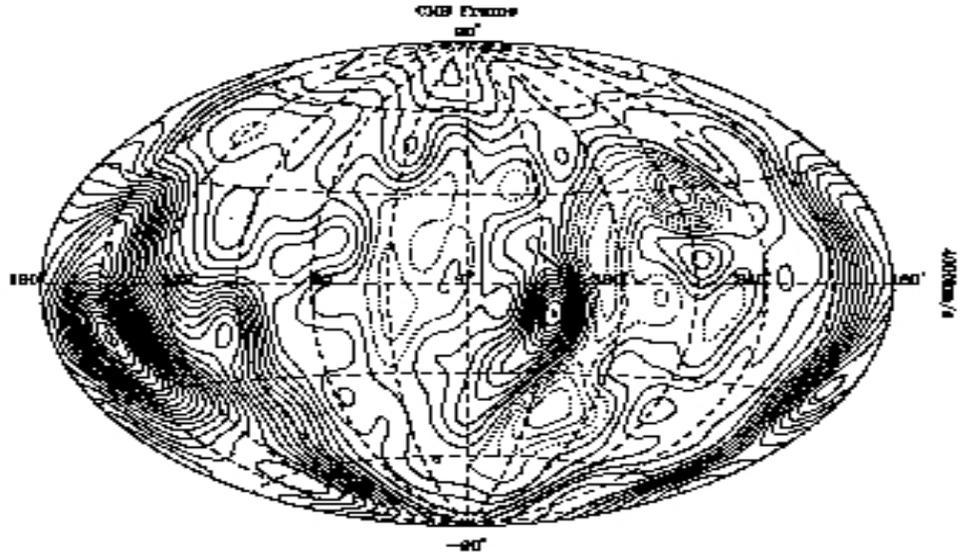,angle=90,width=\textwidth,height=0.4\textheight,clip=} 
\caption[]{Same as Figure~\ref{fig:velshells(a)} but
 evaluated at $4000 \kmps$. This velocity field corresponds with the
 density field shown in Figure~\ref{fig:denshells(b)}. Note
 Perseus-Pisces and N1600 cause the strongest outflow towards
 ($l\simeq150^\circ$, $b\simeq-20^\circ$). There is also a flow
 towards the centre of Pavo-Indus-Telescopium ($l\simeq335^\circ$,
 $b\simeq-25^\circ$) and a weaker one towards Hydra
 ($l\simeq280^\circ$, $b\simeq10^\circ$).}
\label{fig:velshells(b)}
\end{figure*}

\begin{figure*}
\psfig{figure=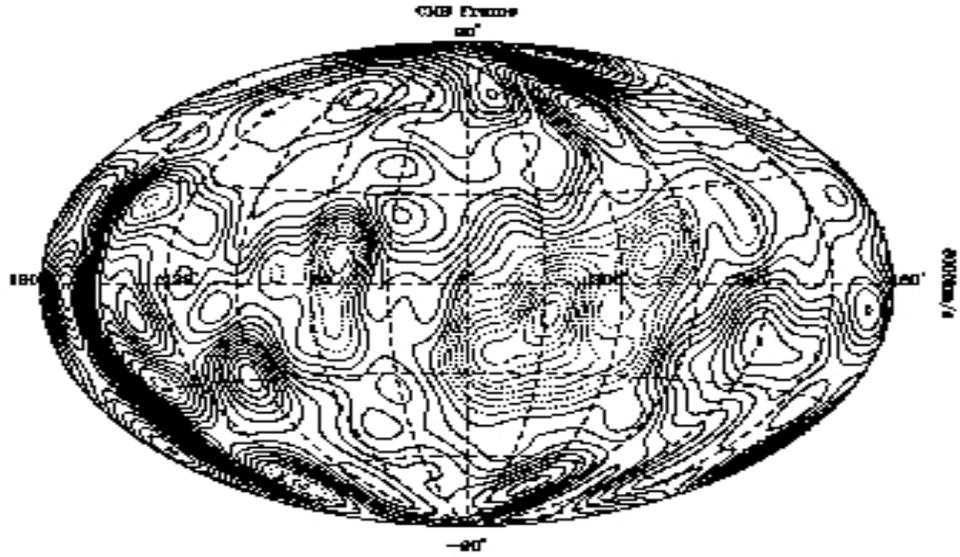,angle=90,width=\textwidth,height=0.4\textheight,clip=} \caption[]{Same as Figure~\ref{fig:velshells(a)} but
 evaluated at $6000 \kmps$. This velocity field corresponds with the
 density field shown in Figure~\ref{fig:denshells(c)}. Note the strong
 outflow towards the Great Wall, especially towards Coma and the
 general velocity flow towards the North Galactic Pole from the South
 Galactic Pole.  There is still some outflow towards Pegasus. Also
 visible is a strong back-side infall towards the Great Attractor
 region and Perseus-Pisces.}
\label{fig:velshells(c)}
\end{figure*}

\begin{figure*}
\psfig{figure=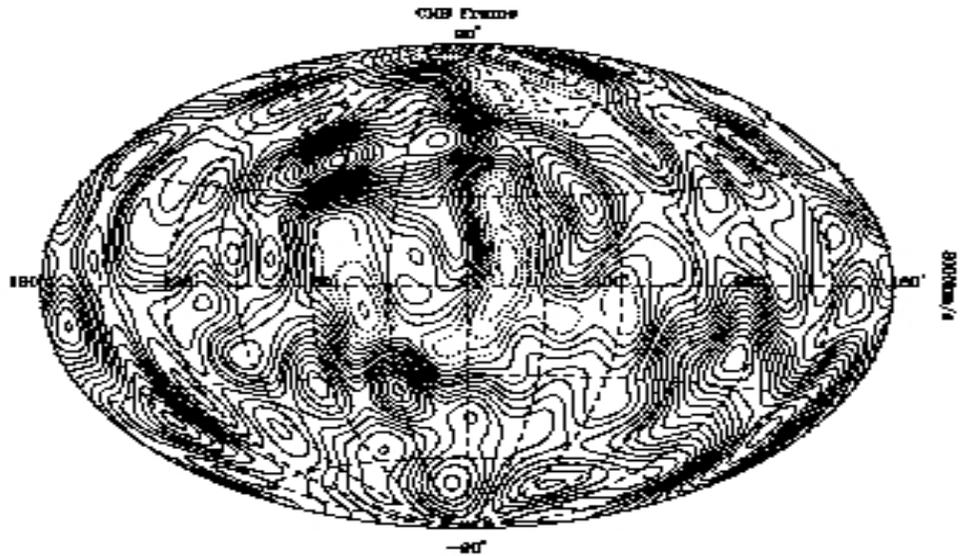,angle=90,width=\textwidth,height=0.4\textheight,clip=} \caption[]{Same as Figure~\ref{fig:velshells(a)} but
 evaluated at $8000 \kmps$. This velocity field corresponds with the
 density field shown in Figure~\ref{fig:denshells(c)}. There is still a
 strong outflow towards the clusters of the Great Wall. The outflow
 towards Shapley ($l\simeq310^\circ$, $b\simeq25^\circ$) is already
 visible.  The Sculptor Wall is implied in the outflow towards the
 Southern Galactic Pole.}
\label{fig:velshells(d)}
\end{figure*}

\begin{figure*}
\psfig{figure=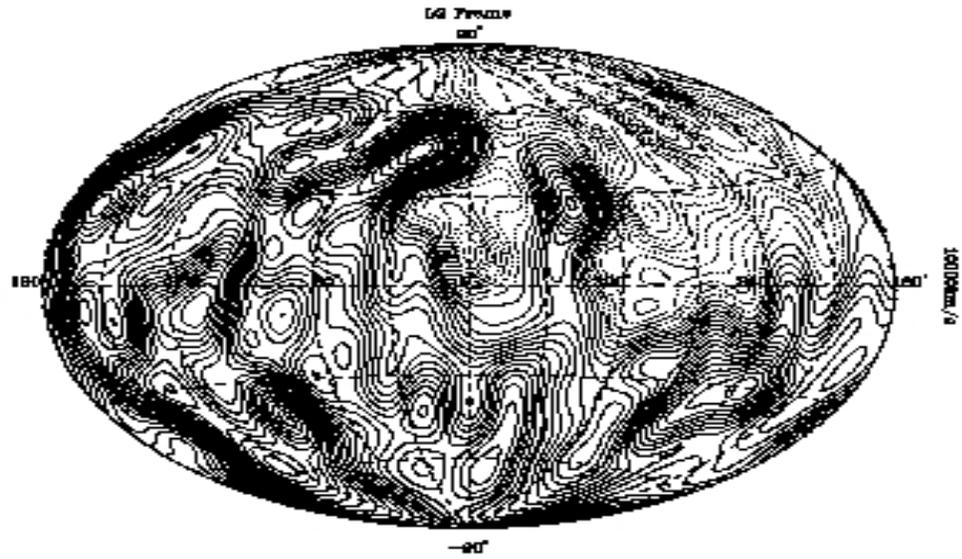,angle=90,width=\textwidth,height=0.4\textheight,clip=} \caption[]{Same as Figure~\ref{fig:velshells(a)} but
 evaluated at $10000 \kmps$ and in the LG Frame instead of the CMB Frame. 
There are three
 superclusters that dominate the outflow, Hercules towards
 ($l\simeq20^\circ$, $b\simeq45^\circ$), Shapley towards
 ($l\simeq320^\circ$, $b\simeq25^\circ$) and Columba towards
 ($l\simeq225^\circ$, $b\simeq-30^\circ$).}
\label{fig:velshells(e)}
\end{figure*}

\begin{figure*}
\psfig{figure=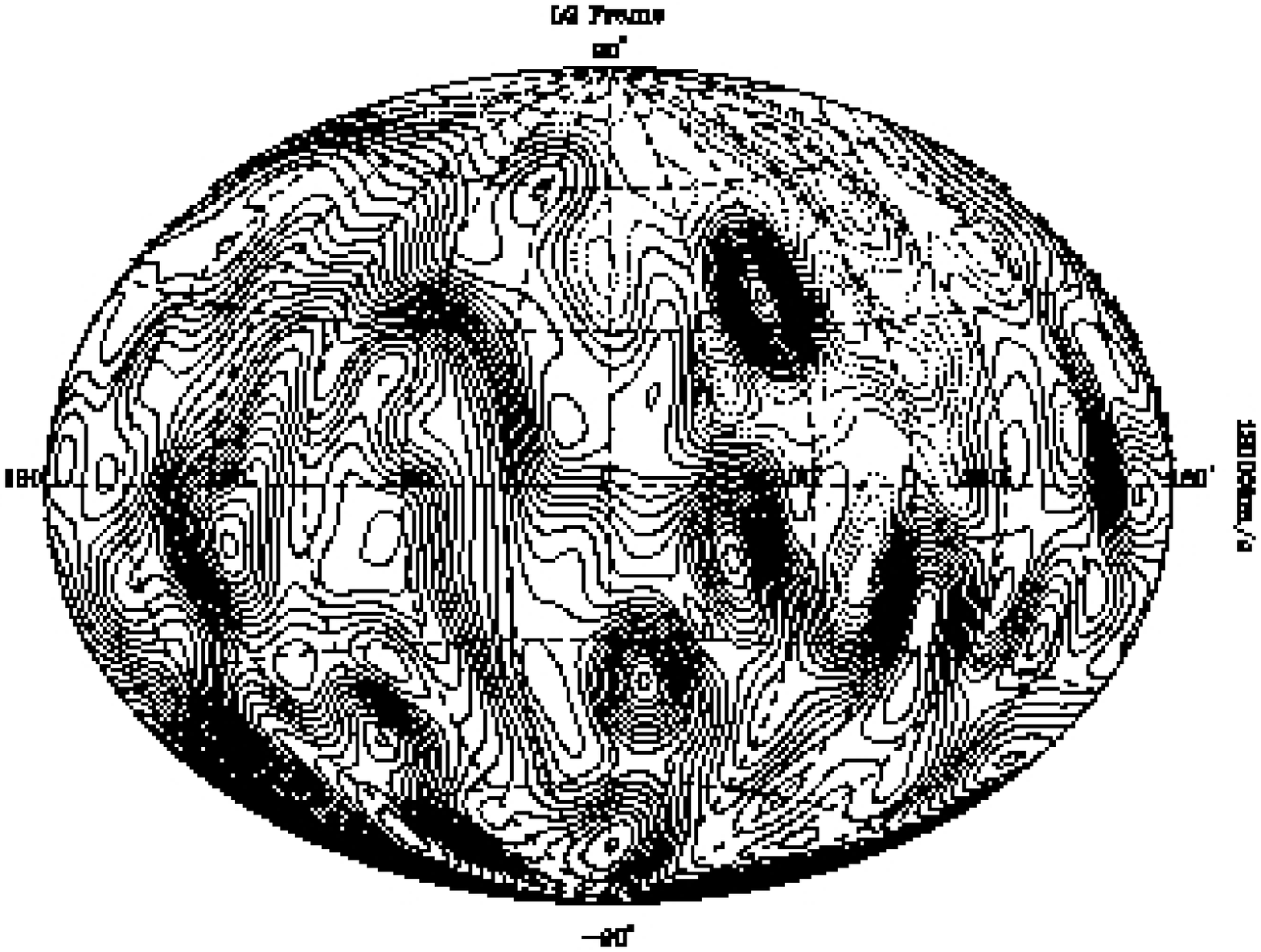,angle=90,width=\textwidth,height=0.4\textheight,clip=} 
\caption[]{Same as Figure~\ref{fig:velshells(a)} but
 evaluated at $12000 \kmps$ and in the LG Frame instead of the CMB Frame.} 
\label{fig:velshells(f)}
\end{figure*}

\begin{figure*}
\psfig{figure=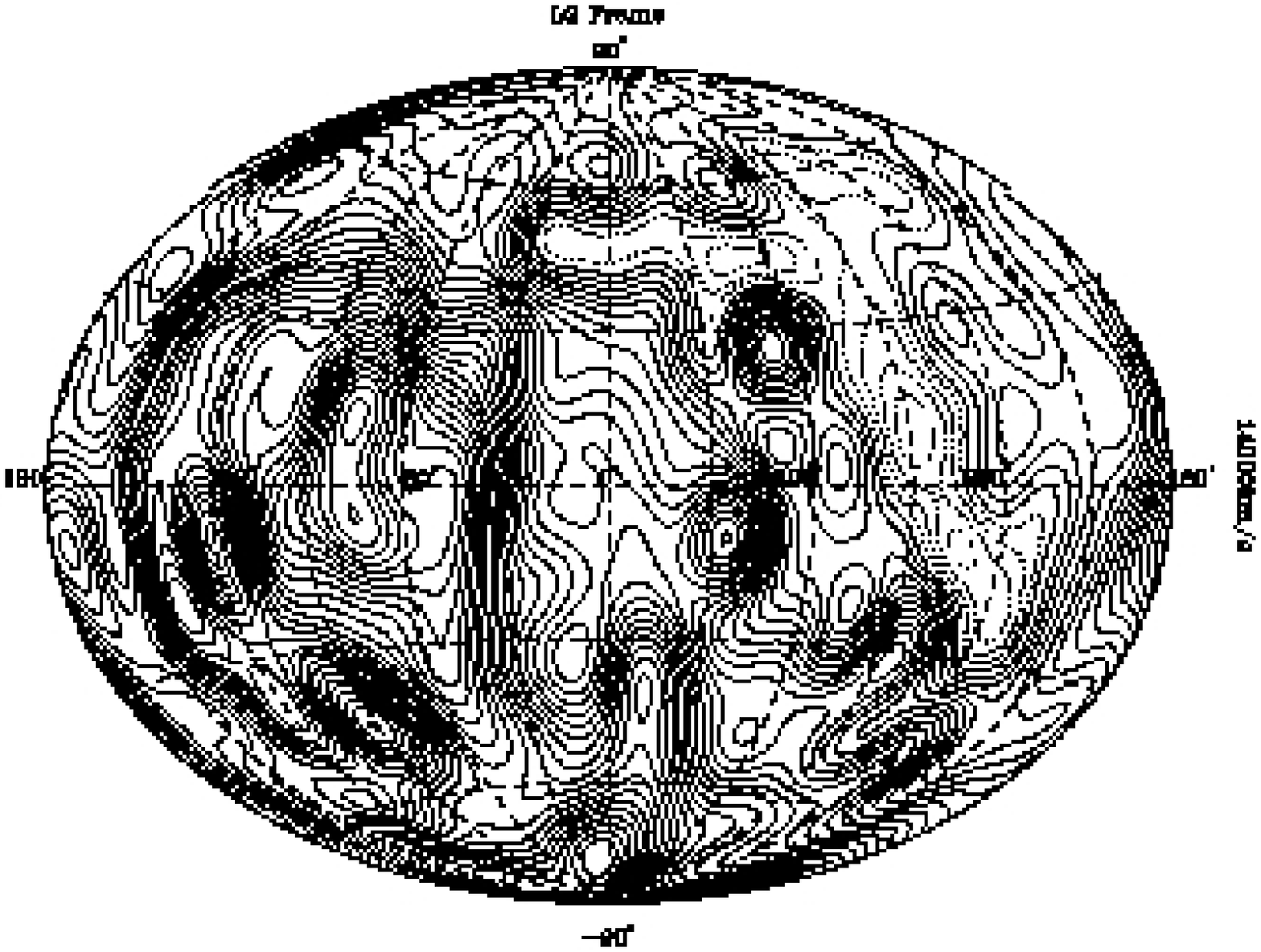,angle=90,width=\textwidth,height=0.4\textheight,clip=} \caption[]{Same as Figure~\ref{fig:velshells(a)} but
 evaluated at $14000 \kmps$ and in the LG Frame instead of the CMB Frame.} 
\label{fig:velshells(g)}
\end{figure*}

\begin{figure*}
\psfig{figure=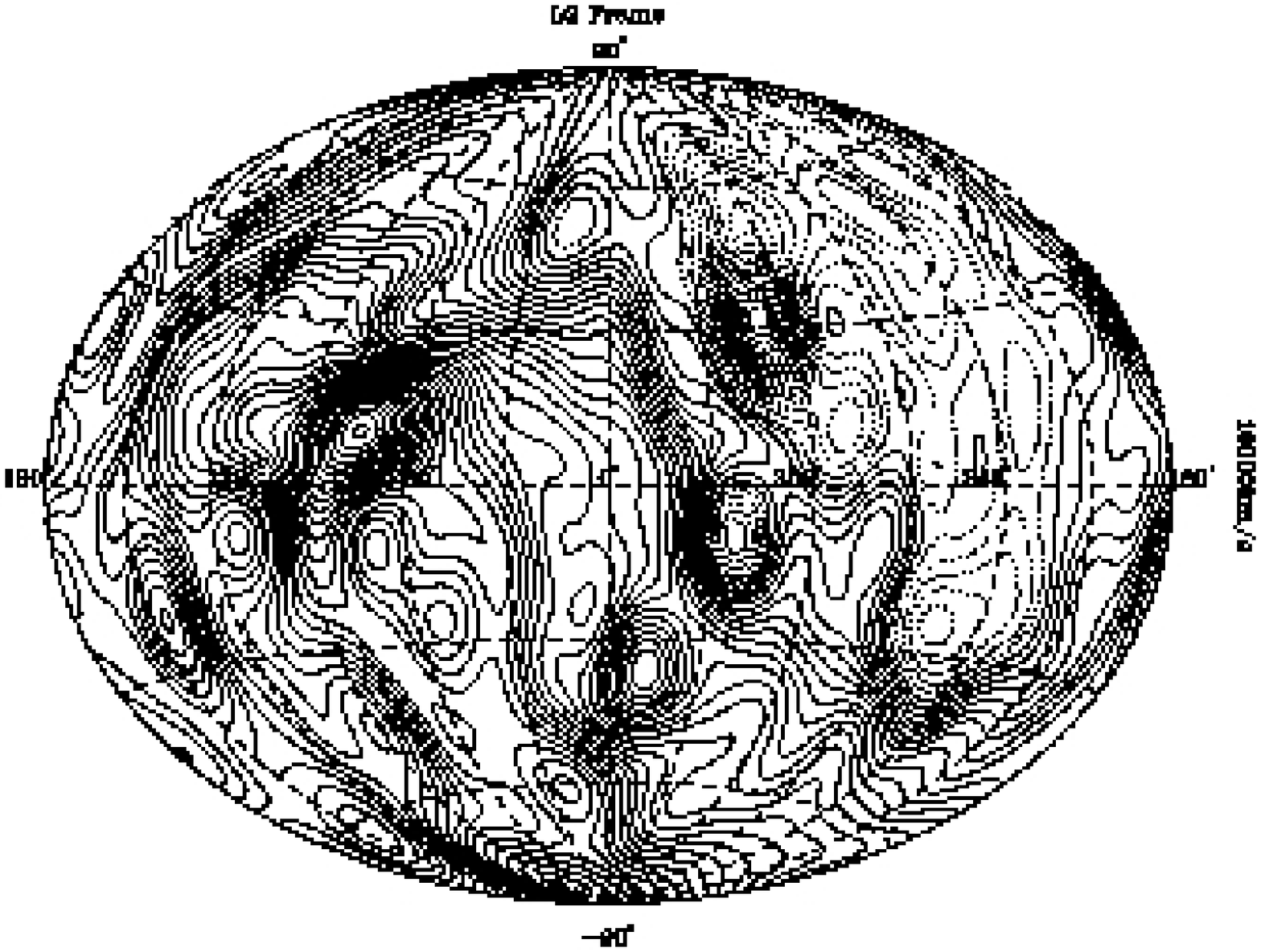,angle=90,width=\textwidth,height=0.4\textheight,clip=} \caption[]{Same as Figure~\ref{fig:velshells(a)} but
 evaluated at $16000 \kmps$ and in the LG Frame instead of the CMB Frame.} 
\label{fig:velshells(h)}
\end{figure*}

Figures~\ref{fig:velshells(a)},~\ref{fig:velshells(b)},~\ref{fig:velshells(c)},~\ref{fig:velshells(d)},~\ref{fig:velshells(e)},~\ref{fig:velshells(f)},~\ref{fig:velshells(g)} and~\ref{fig:velshells(h)} show the Galactic Aitoff projections of the radial velocity fields for $\beta=0.5$ evaluated across the same shells as the
density fields shown in Figures~\ref{fig:denshells(a)},~\ref{fig:denshells(b)},~\ref{fig:denshells(c)},~\ref{fig:denshells(d)},~\ref{fig:denshells(e)},~\ref{fig:denshells(f)},~\ref{fig:denshells(g)} and~\ref{fig:denshells(h)}.
Positive (outflowing) radial velocities are shown as
solid lines and the negative (infalling) radial velocities are shown
as dashed lines. The contour spacing is $|\Delta v_{radial}|=50 \kmps$.

{\bf Figures~\ref{fig:velshells(a)}} \& {\bf ~\ref{fig:velshells(b)}} 
show the CMB frame radial velocity fields at $r=2000
\kmps$ and at $r=4000 \kmps$, respectively. 
The expected scatter in the velocity field is
$\langle\Delta{\bf v}({\bf r})\rangle= 53 \kmps$ at $r=2000
\kmps$ and $\langle\Delta{\bf v}({\bf r})\rangle=62 \kmps$ at $r=4000 \kmps$. 
At $r=2000 \kmps$,
there is a strong outflow towards the Virgo-Great
Attractor ($l\simeq290^\circ$, $b\simeq15^\circ$) and inflow out of the
underdense regions.  Also seen at this distances 
are the effects of Perseus-Pisces
and N1600, causing a strong outflow towards ($l\simeq165^\circ$,
$b\simeq-30^\circ$). 
At $r=4000 \kmps$, Perseus-Pisces and N1600
are the dominant structures causing an outflow towards
($l\simeq150^\circ$, $b\simeq-20^\circ$). There is also a flow towards
the centre of Pavo-Indus-Telescopium ($l\simeq335^\circ$,
$b\simeq-25^\circ$) and a back-side infall to Hydra. 
The back-side infall to the Great
Attractor region becomes strongest at $r=5000 \kmps$ and reaches to 
$v_{\rm infall}=(982\pm400) \beta \kmps$ in the Local Group
frame and $v_{\rm infall}=(127\pm409) \beta \kmps$ in the CMB frame, 
where the quoted values are the mean of the velocity
field around the region and the error bars are the standard deviation from
that mean and the expected scatter in the velocity field added in quadrature. 

Although still visible at  $r=4000 \kmps$, the Great Wall really starts to dominate in
{\bf Figure~\ref{fig:velshells(c)}} at a distance of $r=6000 \kmps$.  At this
distance the velocity scatter rises to 
$\langle\Delta{\bf v}({\bf r})\rangle=72 \kmps$. Here
the fall towards the Coma Cluster is quite apparent at
$l\simeq240^\circ$ and $b\simeq80^\circ$. The velocity flow towards Perseus-Pisces is still strong 
and there is some outflow towards Pegasus
($l\simeq95^\circ$, $b\simeq-60^\circ$), Phoenix ($l\simeq330^\circ$,
$b\simeq-75^\circ$), the Abell clusters A548, A539
($l\simeq230^\circ$, $b\simeq-20^\circ$) and A400 ($l\simeq175^\circ$,
$b\simeq-45^\circ$).  The CMB radial velocities at 
$r=8000 \kmps$, shown in {\bf Figure~\ref{fig:velshells(d)}}, have 
scatter of $\langle\Delta{\bf v}({\bf r})\rangle=87 \kmps$ and 
are mostly outflowing apart from the remnants of back-side infall towards the Great Attractor and the Coma cluster.
By $r=10000 \kmps$, it becomes difficult to compare peculiar velocities in the CMB frame so we switch to the LG frame. 
At $r=10000 \kmps$  
({\bf Figure~\ref{fig:velshells(e)}}, $\langle\Delta{\bf v}({\bf
r})\rangle=108 \kmps$), the dipole velocity of the LG is still visible, however there is an outflow towards
Shapley ($l\simeq310^\circ$, $b\simeq25^\circ$) and the Hercules supercluster 
($l\simeq20^\circ$, $b\simeq45^\circ$).
Shapley outflow becomes stronger at $r=12000 \kmps$ ({\bf
  Figure~\ref{fig:velshells(f)}},$\langle\Delta{\bf v}({\bf r})\rangle=130 
\kmps$), begins to decrease at $r=14000 \kmps$ ({\bf
  Figure~\ref{fig:velshells(g)}},$\langle\Delta{\bf v}({\bf r})\rangle=146 \kmps$)
and at  $r=16000 \kmps$ ({\bf Figure~\ref{fig:velshells(h)}}, $\langle\Delta{\bf v}({\bf r})\rangle=150 \kmps$) there is a strong back-side infall towards it. 

\begin{figure*}
$\begin{array}{cc}
\psfig{figure=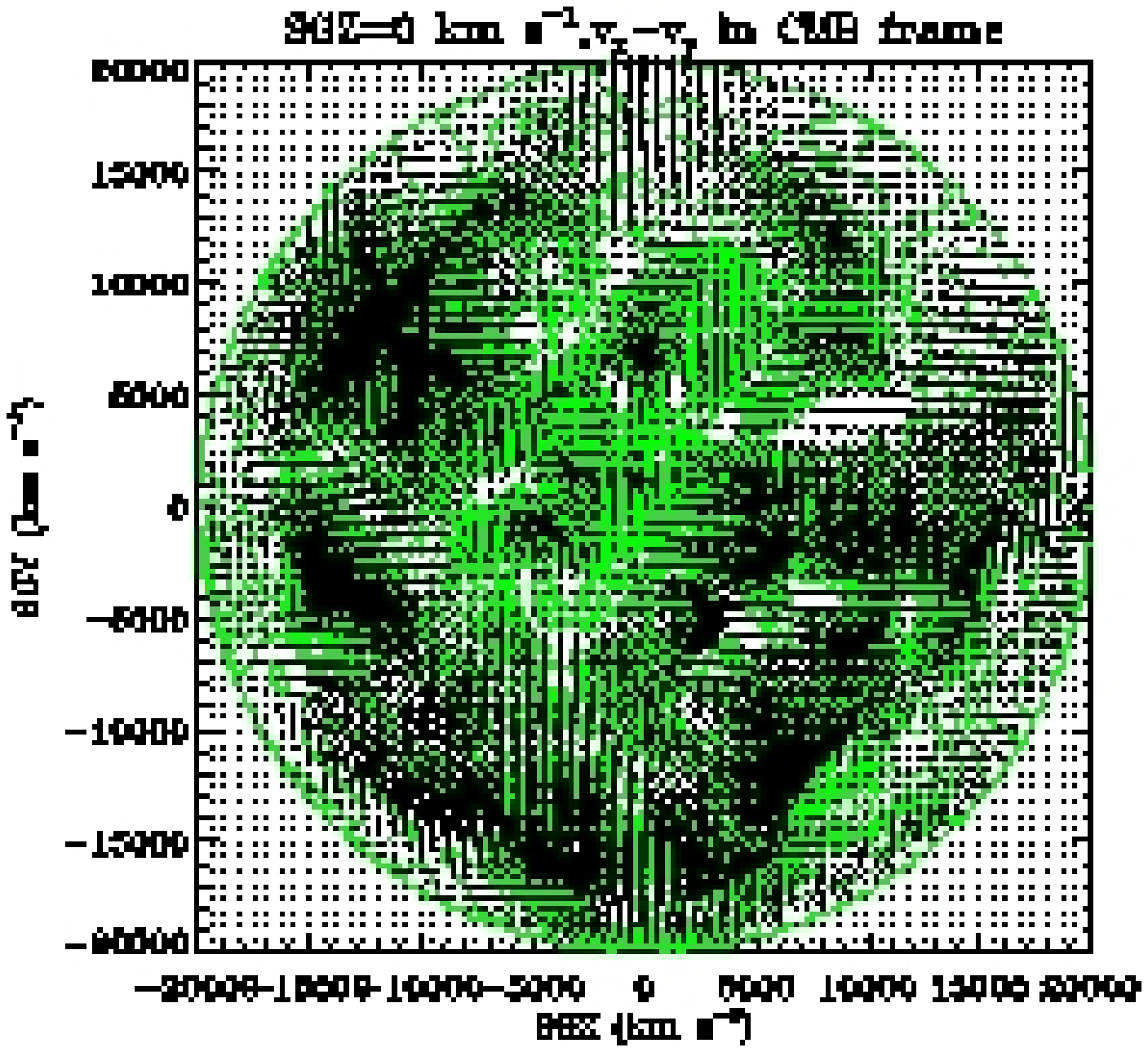,angle=0,height=0.5\textheight,clip=}\\
\psfig{figure=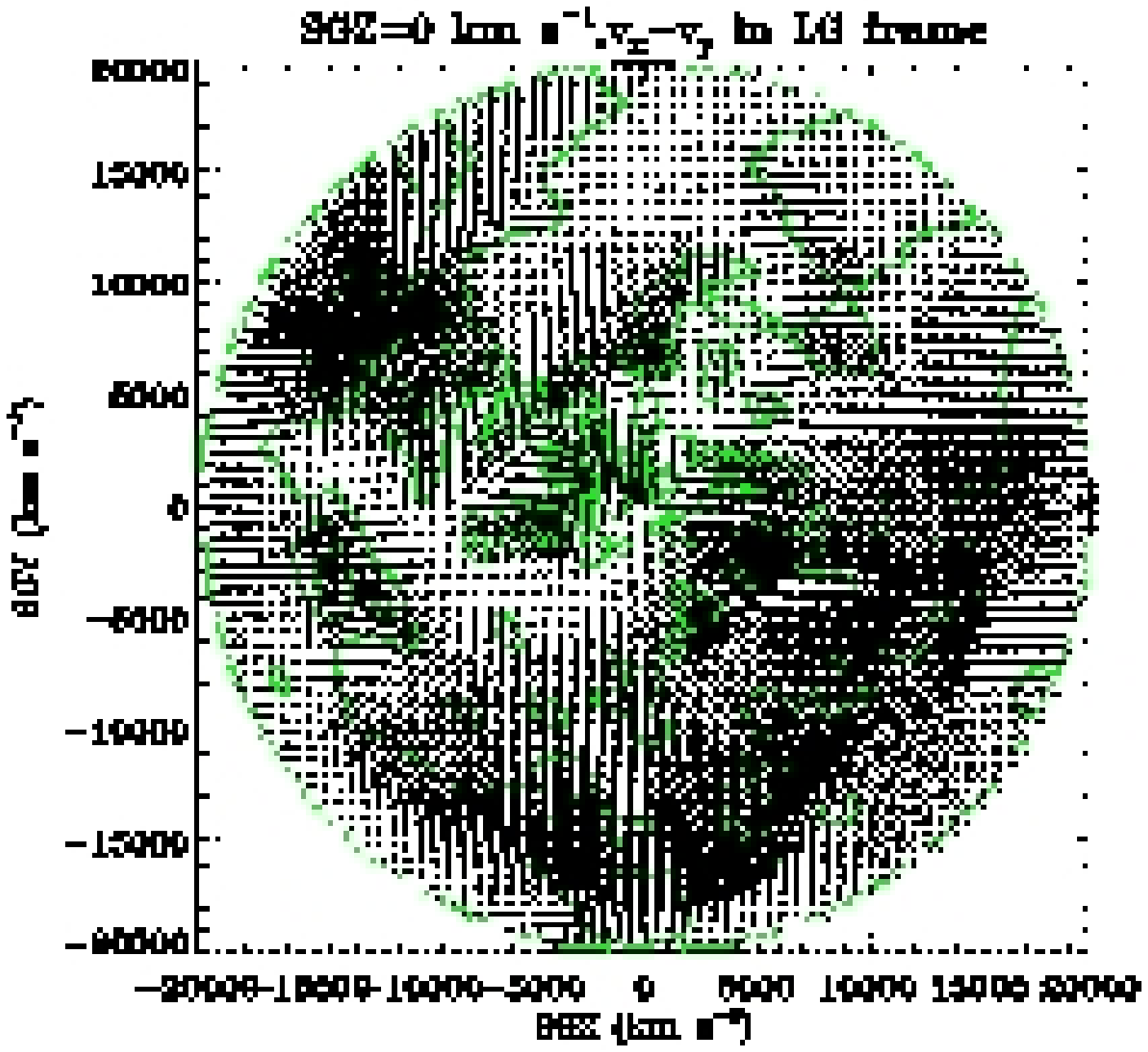,angle=0, height=0.5\textheight,clip=}\\
\end{array}$
\caption[The reconstructed density and velocity fields in the
supergalactic plane] {The reconstructed density fields in the
supergalactic plane. The spacing of the density contours 
is $|\Delta\delta=0.1|$ with dashed
lines denoting the negative and the solid line denoting the positive
density contrast.  Superimposed in black are the reconstructed
three-dimensional peculiar velocity field in the CMB (top plot) 
and the LG (bottom plot) frames. The length of the arrows are about 
$300 \kmps$ per cell. The main overdensities are Hydra-Centaurus (centre-left), Perseus-Pisces (centre-right), Shapley Concentration (upper left), Coma (upper-middle).}
\label{fig:sgplane}
\end{figure*}

{\bf Figure~\ref{fig:sgplane}} shows the 2MRS density and velocity fields in
the Supergalactic plane ($SGZ=0 {\rm kms}^{-1}$) in different
reference frames.  The top plot is the reconstructed
three-dimensional peculiar velocity field in the CMB frame, superimposed
on the reconstructed density field in the Supergalactic plane.  The
bottom plot is the same as the one on the left but the superimposed
peculiar velocity field is reconstructed in the LG frame 
(obtained by subtracting the Local Group velocity ($v_{LG}=627\pm22$
\kmps, towards $l_{LG}=273^\circ\pm3^\circ$,
$b_{LG}=29^\circ\pm3^\circ$).  The
extent of the planar density distribution in the 
Supergalactic plane can clearly be seen in these plots. Since the Local Group is moving 
towards the Great Attractor/Shapley region, the strong outflow towards the positive y-plane
seen in the top plot decreases in the bottom plot. For the same reason, the outflow towards the negative 
y-plane is much stronger in the LG frame (bottom plot) than in the CMB frame. 
The most
intriguing result is the connectivity of the Shapley supercluster
(SGX,SGY) = (-12000, 7000) to the Great Attractor region. The Virgo
cluster at $(SGX,SGY) = (0, 1000)$ also appears connected to the Great
Attractor. The Perseus-Pisces supercluster also is visible at
(SGX,SGY) = (2500,-4500) and Camelopardalis is at (SGX,SGY) = (45,
-30). The other visible superclusters are Columba at (SGX,SGY) = (7500,
-11000) and Hercules (SGX,SGY)=(-13000, -4000).  In both plots, the
outflow is dominated by Shapley, particularly in the CMB frame.  
The flow
towards Virgo and the Perseus-Pisces region is also apparent and the
flow towards the Great Attractor is clearly visible.  Also observed is
a significant flow towards Columba and some towards Hercules. 
After 16000 $\kmps$, the
shot noise is too high and the density field becomes non-uniform.

\begin{figure}
\psfig{figure=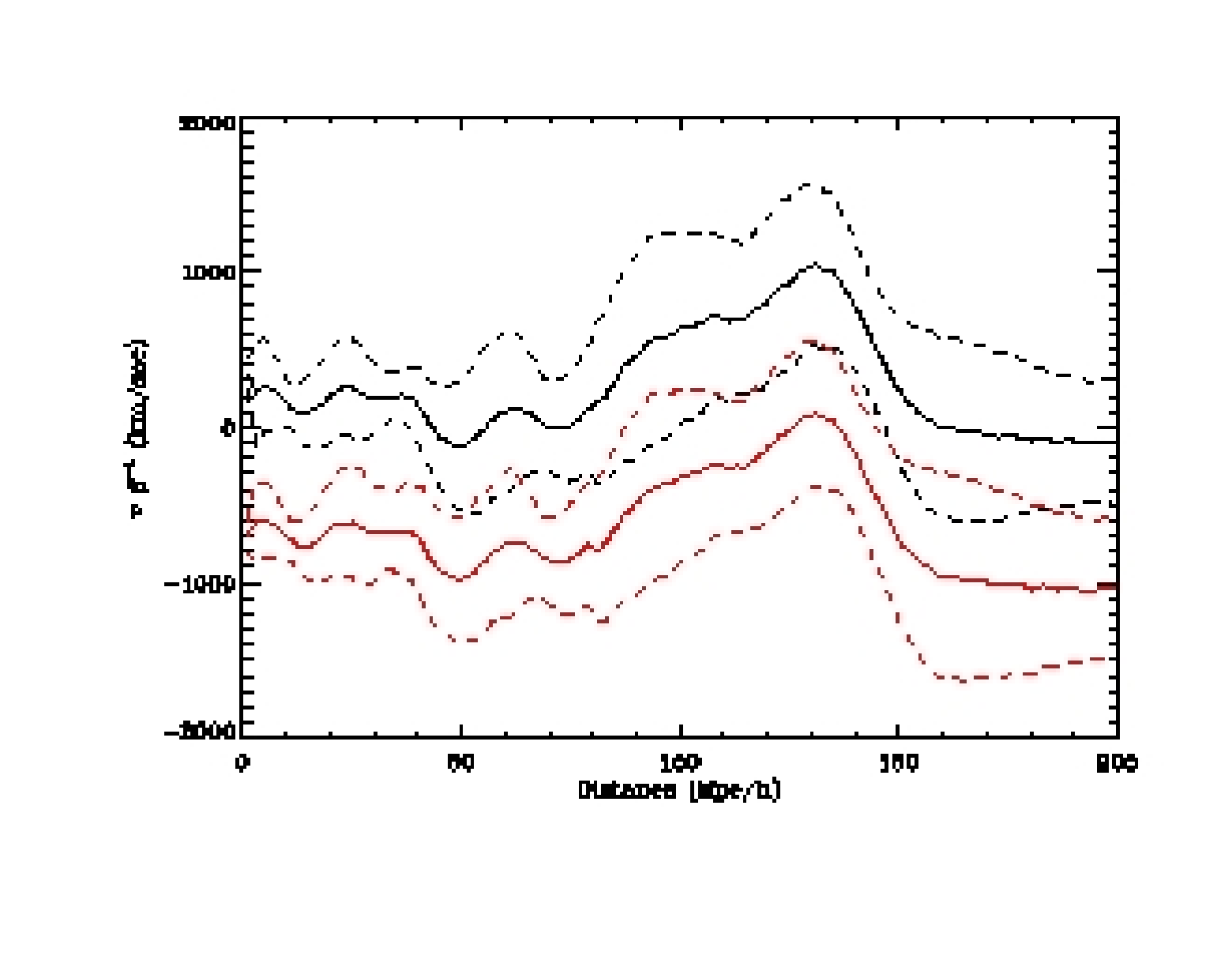,angle=0,width=0.5\textwidth, clip=}
\caption[]{The radial velocity field in the region of the
GA and the Shapley Superclusters as a function of distance. The red lines are
for the LG and the black lines are for the CMB frame. The solid lines denote
the mean of the reconstructed velocities and the dashed lines are the errors
calculated from the standard deviation of velocities and the expected velocity
scatter added in quadrature. The regions with negative velocities move towards us and positive velocities move away from us.}
\label{fig:infall}
\end{figure}

Over the
past years, there has been some disagreement over the existence of
back-side infall towards the GA (e.g. Dressler \& Faber 1990 and
Mathewson \etal 1992).  We find a clear back-side infall, visible in both plots and in particular the one in the LG frame. 
The highest velocity back-falling velocity
around the GA region is $800\pm 600 \beta^{-1} \kmps$ (at $~55 h^{-1}$ Mpc) 
in the CMB frame.
In {\bf Figure~\ref{fig:infall}}, we plot the average radial velocity field in towards the centres of the 
GA ($290^\circ \leq l \leq 320^\circ$, $-25^\circ \leq b \leq 60^\circ$$-25^\circ \leq b \leq 70^\circ$, $r\leq80 h^{-1}$ Mpc ) and the Shapley ($290^\circ \leq l \leq 320^\circ$, $40^\circ \leq b \leq 70^\circ$, $r\geq80 h^{-1}$ Mpc) Superclusters as a function of distance. The red lines are
for the LG frame and the black lines are for the CMB frame. The solid lines denote
the mean of the reconstructed velocities and the dashed lines are the errors
calculated from the standard deviation of velocities in that region and the expected velocity
scatter added in quadrature. In this plot, the back infall onto the GA and the
subsequent outflow towards Shapley are clearly visible. An observer in the GA would see the surrounding nearby galaxies moving towards her. On larger scales, she would observe a dipole velocity towards Shapley. If she was in the Shapley Supercluster (at $r\approx130 h^{-1}$ Mpc), she would 
observe that the surrounding galaxies are moving towards her.   
Whilst, this back-side infall onto the GA was also observed by FLHLZ, WLF and
Schmoldt \etal (1999), 
other authors who applied
different reconstruction techniques to $IRAS$ galaxy catalogues (e.g. Davis, Nusser \& Willick
1996; Branchini \etal 1999; Valentine, Saunders and Taylor 2000) find little evidence of it. 
Although the 2MRS sample is the densest all-sky sample to date and as such
allows the most precise reconstruction of the local velocity field, our results
are still not conclusive. In a forthcoming paper we will compare the
reconstructed fields with the observed velocity fields based on the distance measurements to elliptical galaxies in the  
6dFGS and a new survey of spirals in the I Band (SFI++) to assess the significance of this
back-side infall.

\subsection{Alternative Reconstruction Methods}
\begin{figure*}
$\begin{array}{cc}
\psfig{figure=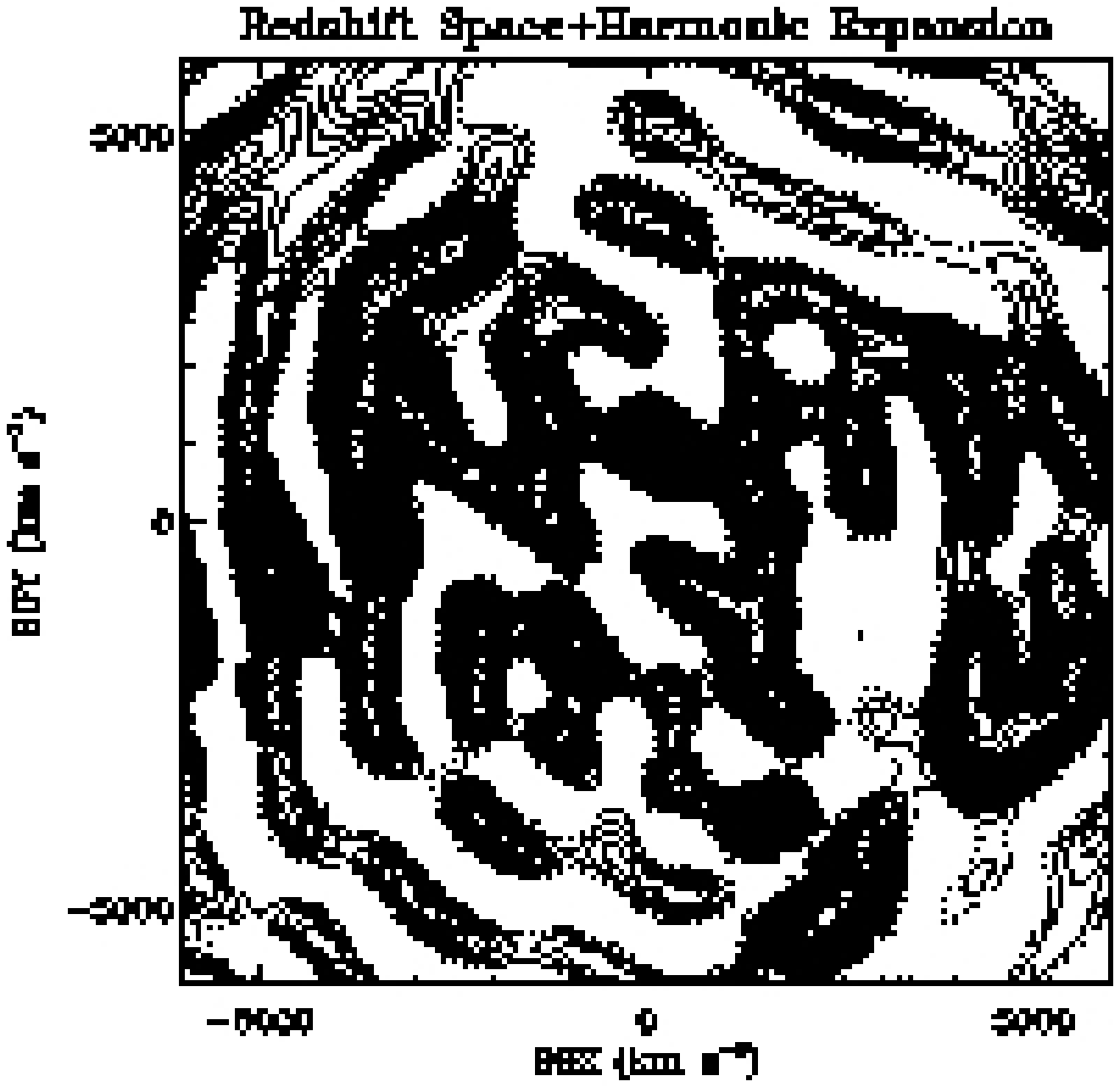,angle=0,width=0.5\textwidth, clip=}&
\psfig{figure=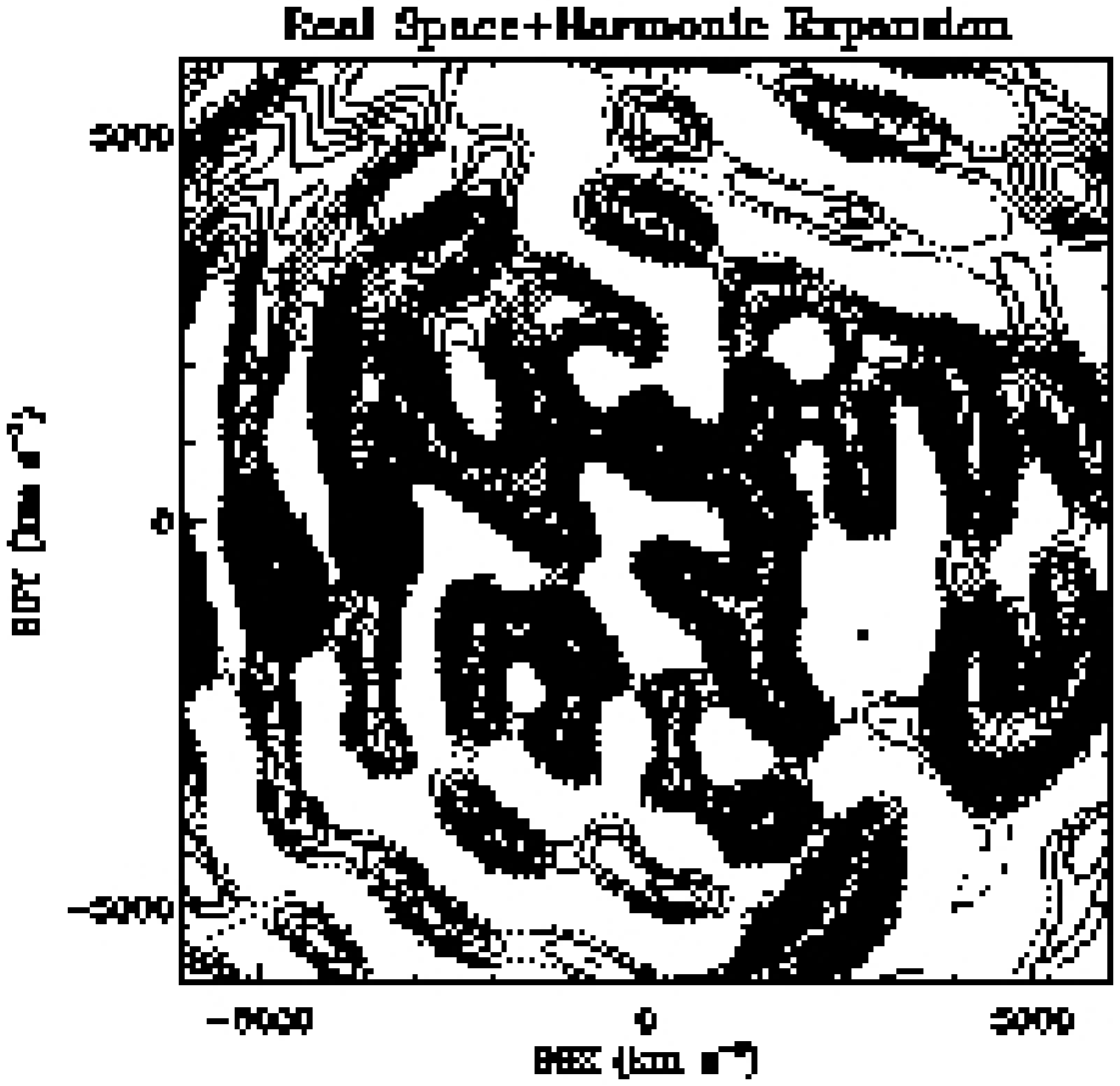,angle=0,width=0.5\textwidth, clip=}\\
\psfig{figure=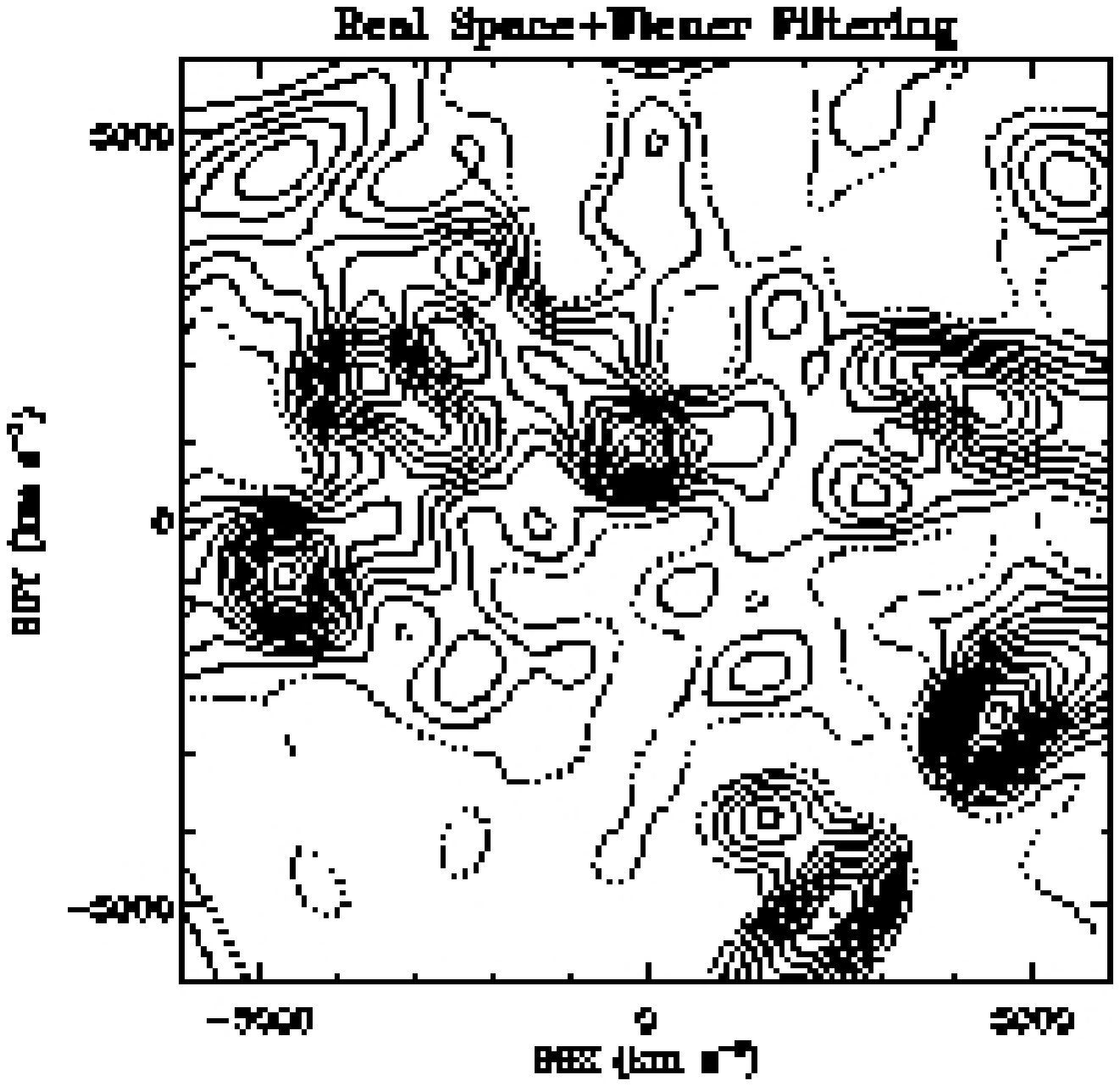,angle=0, width=0.5\textwidth, clip=}&
\psfig{figure=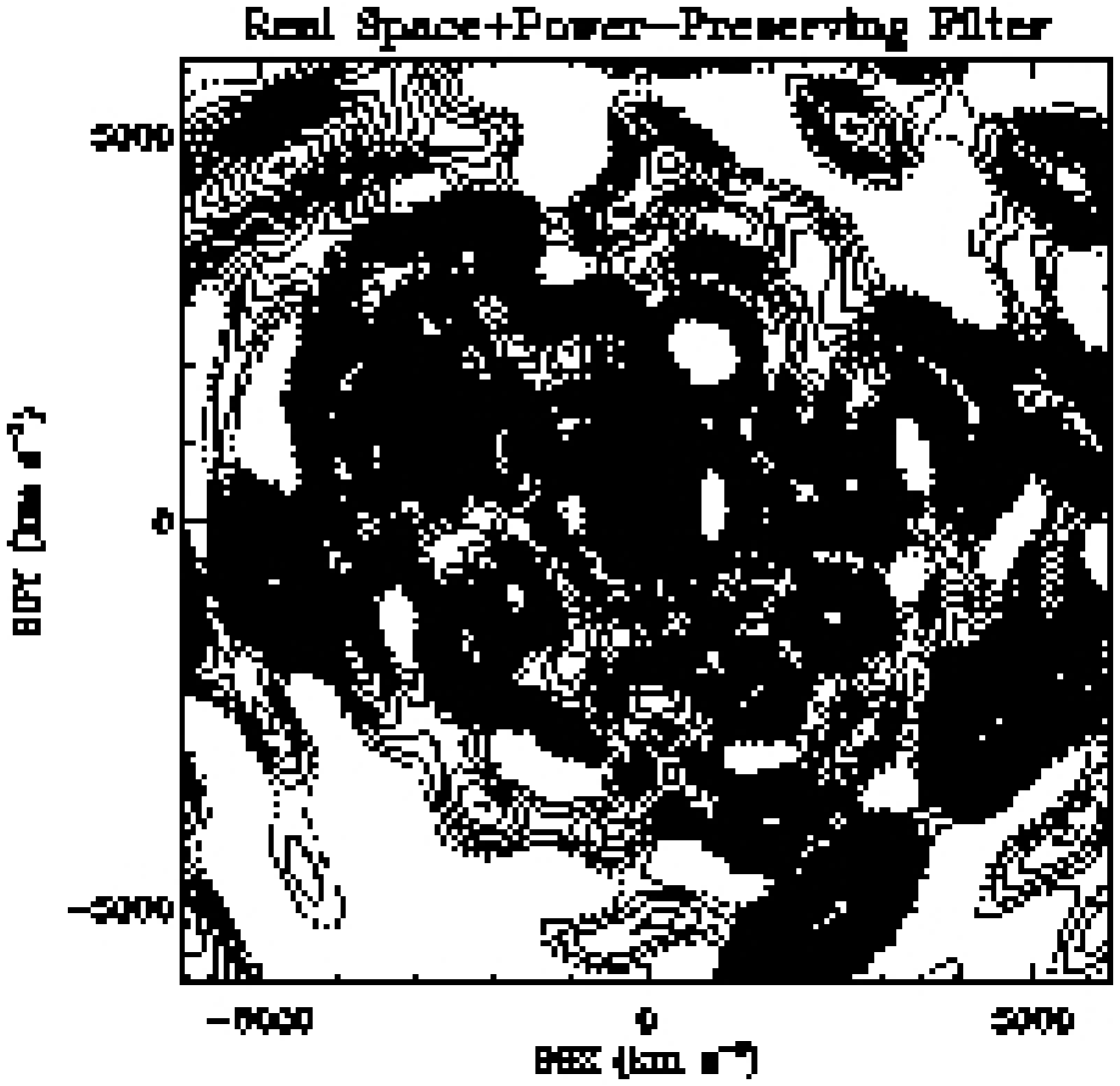,angle=0,width=0.5\textwidth, clip=}\\
\end{array}$
\caption[]{Recontructions of the 2MRS density field in the Supergalactic
  Plane. All contours are spaced at $|\Delta \delta| = 0.5$, with solid (dashed)
  contours denoting positive (negative) contours. Top left: The density field
  expanded in spherical harmonics with no smoothing. Top right: Same as in top
  left but with the harmonics corrected for redshift distortion. Bottom left:
  Same as in top right but smoothed by the Wiener Filter. Bottom Right: The
  real space density field smoothed by the power preseving filter.}
\label{fig:comp_wf}
\end{figure*}

In order to assess the
effects of the systematic errors of our method, we also carried out 
the reconstructions using alternative methods. {\bf Figure~\ref{fig:comp_wf}} shows
several of these reconstructions on the Supergalactic Plane (SGP) out to 
$6000 \kmps$. The top left
plot is the unsmoothened harmonic reconstruction of the 2MRS density field in
redshift space. It can clearly be seen that 
the harmonics are strongly affected by shot noise. 
Overdensities follow circular paths about the origin as a result
of correlations introduced by noise. These artifacts
are still clearly evident in the top right plot where the density field is
corrected for redshift distortions (using Equation~\ref{eq:rhostorhor}) 
but without noise supression. The correction of the redshift distortions
reduces the overall amplitude of the density contrasts and makes the
structures more spherical. This reconstruction corresponds to the unbiased minimal variance (umv) estimation of Zaroubi (2002) since in our case the redshift distortion matrix is invertible. 
The bottom left panel shows the real space density
field after the Wiener Filter has been applied. As expected the Wiener Filter 
mitigates the effects of shot noise, removing the artifacts and high frequency
harmonics from the map, however, the Wiener 
reconstructed field is biased towards the
mean field which is zero by construction. In fact, the variance of the Wiener Filtered field is always smaller than the variance of the {\it true} underlying field. 
In order to overcome this drawback, Yahil (1994) introduced an
alternative filter which preserves the power in the underlying field. This filter is defined as the square root of the Wiener Filter. With this definition, the filtered field equals to that of the true field.   
The bottom right panel shows the real-space density field reconstructed using
Yahil's filter. Whilst the estimated field is unbiased, the reconstruction is not
optimal in the sense of minimum variance and there are artifacts due to
insufficient reduction of noise. 

\subsection{The acceleration of the Local Group}\label{subsec:2mass:LGdipole}
A detailed calculation of the acceleration of LG using
the 2MRS data was given by Erdo\u{g}du \etal (2006). In this section, the LG velocity is reconstructed from the density field. 
There are two main advantages of the analysis
performed in this paper. Firstly, working with spherical harmonic
allows for reliable correction of the velocity flows and
secondly, the Wiener filtering suppresses the noise self-consistently.
The LG dipole is linear
in the galaxy density field (assuming linear biasing). Consequently,
the dipole computed from the Wiener filter algorithm will be an
optimal (in the sense of minimum variance) estimate of the dipole due
to matter within the reconstruction volume as long as non-linear
effects can be neglected. 

\begin{figure*}
\psfig{figure=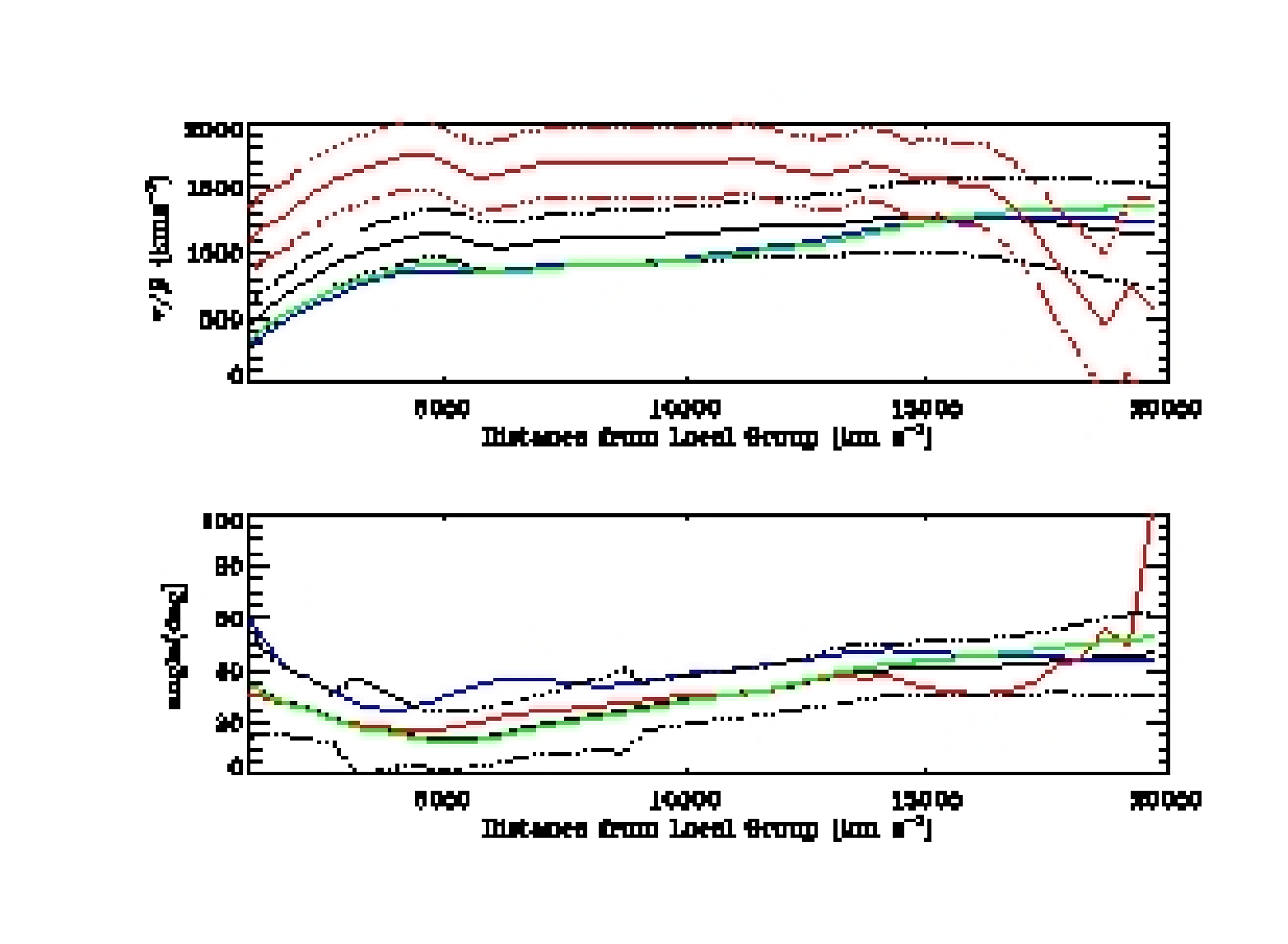,angle=0,width=\textwidth,clip=}
\caption[The amplitude of reconstructed Local Group velocity and the
it's angle to CMB dipole]{Top Plot: The growth of acceleration of the
Local Group due to galaxies within a series of successively larger
concentric spheres centred on the local group. The growth of estimated
shot noise is also shown as dashed lines.
The black line is the dipole reconstructed with $\beta=0.5$. 
The blue line is the dipole reconstructed with $\beta=0.5$ 
from the CMB redshifts. 
The green 
line is the dipole reconstructed with $\beta=1.0$. 
The red line is the dipole from Erdo\u{g}du \etal (2006) and 
the dashed line is the associated shot noise. 
Bottom: The angle between the CMB dipole
($l=273^\circ$, $b=29^\circ$) and the reconstructed Local Group
velocity. The lines are colour coded to coincide with the top plot. 
The dashed lines denote the shot noise errors.}
\label{fig:dipolewiener}
\end{figure*}

The reconstructed 2MRS dipole (Equation~\ref{eq:v(r)LGsph}) and the
expected scatter (Equation~\ref{eq:vscatter}) in the reconstruction is plotted as a function of distance 
in the top plot of Figure~\ref{fig:dipolewiener} . The bottom plot of
Figure~\ref{fig:dipolewiener} shows the angle between the direction of
the LG dipole and the CMB dipole ($l=273^\circ$,
$b=29^\circ$).  
The black line is the dipole reconstructed with $\beta=0.5$ (rescaled 
to $\beta=1.0$) and the dashed line 
denotes the scatter in the velocity reconstruction. The inversion of the radial coupling matrix introduces a non-linear relationship between the value
of $\beta$ and the amplitude of the reconstructed acceleration.  To
illustrate this, we also plot the dipole reconstructed with $\beta=1.0$ (the green line). The amplitudes of the dipole reconstructions using 
$\beta=0.5$ and $\beta=1.0$ are in good agreement (within the expected scatter) 
and the misalignment angles are almost the same up to 13000 \kmps.
The blue line is the dipole reconstructed with $\beta=0.5$ 
from the CMB redshifts. As expected, the amplitude of the 
dipole is underestimated at nearby distances and the misalignment angle is bigger than the dipole reconstructed from the LG redshifts. We conclude that, since the LG dipole is dominated by nearby galaxies, it is more accurate to use the LG redshifts in the reconstruction.

We also plot the number-weighted dipole obtained using raw redshifts 
(solid red line Erdo\u{g}du \etal 2006) and the associated shot noise (dashed red line). 
The bottom plot indicates that the 
direction of the Wiener-reconstructed dipole is consistent with the dipole obtained 
in Erdo\u{g}du \etal 2006).  
Both dipoles show
the `tug of war' between the Great Attractor and
Perseus-Pisce and 
the dipole velocities are dominated by structure within the distance of 50
$h^{-1}$ Mpc. However, 
the amplitude of the Wiener dipole is less than the number-weighted dipole. 
This is due to the fact that in the previous paper, we did not filter the shot noise errors. Furthermore, in 
the previous paper,  
we did not take peculiar velocities into account when calculating the number-weighted dipole 
and these will lead to the over-estimation of the LG-frame dipole (the {\it Rocket
  Effect}). 
The value for $\beta$ we obtain by equating the CMB dipole measurements with our reconstructed dipole reflects the decrease in the amplitude. We get 
$\beta=0.54\pm0.12$ at 13000 \kmps whereas we find $\beta=0.40\pm0.09$ at 13000 \kmps in Erdo\u{g}du \etal (2006). 
The convergence of the
direction of the LG dipole does not seem affected by the velocity
corrections but there is in general better alignment between the
Wiener-reconstructed LG dipole and the CMB dipole than the
LG dipole, calculated directly from the raw data and the CMB
dipole.  The misalignment angle
of the Wiener dipole from the LG redshifts is almost as that of the raw-data dipole 
up to 13000 \kmps and the angle is $39^\circ$ at this distance. 
At least half of this misalignment is probably 
due to the fact that when we modeled the LG dipole, 
we assumed all the galaxies have the same weight (see Erdo\u{g}du \etal 2006).
Also our analysis uses linear perturbation theory which is correct only to
first order $\delta$.  There may be contributions to the LG
dipole from small scales which would cause gravity and the velocity
vectors to misalign.

\section{Summary and Conclusions}\label{sec:2mass:conc} 
In this paper, the Wiener reconstruction technique was applied to
recover real-space density and velocity fields from 2MRS. The
observed density field was first expanded in terms of spherical
harmonics and spherical Bessel functions (Equations~\ref{eq:denexp}
and~\ref{eq:deltalmn}) which are both orthogonal and which together
satisfy Poisson's equation. Then, the redshift distortions,
described by the coupling matrix ${\bf Z}$,
were corrected for by the matrix inversion of ${\bf Z}$ in
Equation~\ref{eq:rhostorhor}, assuming linear theory and a value of
$\beta=\Omega^{0.6}/b=0.5$. After the inversion, the real-space
density field was estimated using the Wiener filter for a given CDM
power spectrum with $\sigma_8 = 0.7$ and $\Gamma =0.2$
(Equations~\ref{eq:rholmnwf},~\ref{eq:S} and~\ref{eq:Nlmn}). Finally,
the velocity field is reconstructed from the Wiener-filtered density
field using Equations~\ref{eq:vlmn}.

The reconstructed density field resolves most of the known structures as
well as new clusters and voids.  
The density and velocity fields agree well with
the $IRAS$ PSCz and 1.2-Jy fields derived using the Wiener filter 
technique. There is a back-side infall towards the Great Attractor region,
in agreement with Schmoldt \etal (1999) and FLHLZ. There is a
strong outflow towards the Shapley supercluster which suggests that
Shapley plays an important dynamical role in the local velocity
field.

The LG dipole derived from the reconstructed velocity field
is lower in amplitude at
all distances than the LG dipole derived from the raw
data by Erdo\u{g}du \etal (2006).  The misalignment angle between the CMB
dipole and the 2MRS dipole decreases down to $13^\circ$ at 5000 \kmps and its amplitude is about 800 $\beta \kmps$. 
If our canonical value for $\beta$ is correct then this suggests 
most of the
acceleration is generated within 5000 \kmps and the structure at
higher distances contributes less than $30\%$ to the LG
dipole. The `tug-of-war' between the Great Attractor and the
Perseus-Pisces supercluster to which the dipole is commonly attributed
can clearly be seen in Figure~\ref{fig:velshells(a)}.  The 2MRS dipole
agrees well with the best-fit value for the canonical
model to the data with $\beta=0.5$, for fixed $\Gamma =0.2$
and $\sigma_8 = 0.7$.

The Wiener filter technique performs very well and provides a rigourous reconstruction method  
determined by the data and the expected signal.  
In a forthcoming paper, we will compare the reconstructed fields with the peculiar velocities from the 6dF survey. 
This will allow us to test the validity of our reconstructions and enable the refinement of our method.

\section*{ACKNOWLEDGEMENTS}
Special thanks to Peter Coles, Yehuda Hoffman, Michelle Lanyon, Karen Masters
and Chris Short for their valuable comments.
PE would like thank the University College London for its hospitality
during the completion of this work. OL acknowledges a PPARC Senior
Research Fellowship.  JPH, LM, CSK, NM, and TJ are supported by NSF
grant AST-0406906, and EF's research is partially supported by the Smithsonian Institution.  
DHJ is supported as a Research Associate by
Australian Research Council Discovery-Projects Grant (DP-0208876),
administered by the Australian National University.  This publication
makes use of data products from the Two Micron All Sky Survey, which
is a joint project of the University of Massachusetts and the Infrared
Processing and Analysis Center/California Institute of Technology,
funded by the National Aeronautics and Space Administration and the
National Science Foundation. This research has also made use of the NASA/IPAC Extragalactic Database (NED) 
which is operated by the Jet Propulsion Laboratory, California Institute of Technology, under contract 
with the National Aeronautics and Space Administration and the SIMBAD database,
operated at CDS, Strasbourg, France.

{}

\end{document}